\documentclass[12pt]{article}
\usepackage{amssymb,amsmath} 
\usepackage{graphicx}
\usepackage{epsfig} 
\usepackage{latexsym}
\usepackage{psfrag}   
\usepackage{subfig}  
\usepackage{enumitem}
\usepackage{caption}
\usepackage{booktabs}  
\usepackage{tabstackengine}
\usepackage{braket}
\usepackage{mathtools}
\usepackage{protosem}
\usepackage{changepage}
\usepackage{wasysym}
\usepackage[numbers,sort&compress]{natbib}
\usepackage{textcomp}
\usepackage[utf8]{inputenc}
\usepackage{pifont}
\usepackage{bbm} 
\usepackage[colorlinks=true,      linkcolor=blue,      urlcolor=blue,      
            filecolor=blue,      citecolor=blue,       pdfstartview=FitH,     
						pdfpagemode=UseNone,      bookmarksopen=true]{hyperref} 

\usepackage{tikz}
\usepackage{mathrsfs}
\usepackage[inner=2.9cm,outer=2cm]{geometry}
\usepackage[toc]{appendix}
\usepackage{color,soul} 
\usepackage{datetime}

\usepackage{setspace}
\definecolor{darkblue}{rgb}{0.2, 0, 0.8}
\definecolor{darkgreen}{rgb}{0.2, 0.71, 0}

\numberwithin{equation}{section}

\newcommand{\bea}{\begin{eqnarray}}
\newcommand{\eea}{\end{eqnarray}}
\newcommand{\ba}{\begin{eqnarray}}
\newcommand{\ea}{\end{eqnarray}}

\newcommand{\beq}{\begin{equation}}
\newcommand{\eeq}{\end{equation} }
\newcommand{\beqa}{\begin{eqnarray}}
\newcommand{\eeqa}{\end{eqnarray}}
\newcommand{\beqar}{\begin{eqnarray*}}
\newcommand{\eeqar}{\end{eqnarray*}}

\definecolor{cardinal}{rgb}{0.6,0,0}
\definecolor{darkgreen}{rgb}{0,0.4,0}
\definecolor{purple}{rgb}{0.5, 0, 0.5}
\definecolor{golden}{rgb}{0.92, 0.7, 0}
\definecolor{midnight}{rgb}{0, 0, 0.5}
\definecolor{darkblue}{rgb}{0, 0, 0.7}

\begin{document}  


\phantom{AAA}
\vspace{-10mm}

\begin{flushright}

IPHT-T18/124\\

\end{flushright}

\vspace{16mm}

\begin{center}

{\huge {\bf Bubbling the NHEK}}

\bigskip

\vspace{20mm}

{\large
\textsc{ Pierre Heidmann}}

\vspace{15mm}

Institut de Physique Th\'eorique,\\
Universit\'e Paris Saclay,\\
CEA, CNRS, F-91191 Gif sur Yvette, France \\

\vspace{11mm} 
{\footnotesize\upshape\ttfamily pierre.heidmann @ ipht.fr} \\

\vspace{28mm}
 
\textsc{Abstract}

\end{center}

\begin{adjustwidth}{6mm}{6mm} 
 
\vspace{1mm}
\noindent
We build the first family of smooth bubbling microstate geometries that are asymptotic to the near-horizon region of extremal five-dimensional Kerr black holes (NHEK). These black holes arise as extremal non-supersymmetric highly-rotating D1-D5-P solutions in type IIB string theory on T$^4\times$S$^1$. Our solutions are asymptotically NHEK in the UV and end in the IR with a smooth cap. In the context of the Kerr/CFT correspondence, these bubbling geometries are dual to pure states of the 1+1 dimensional chiral conformal field theory dual to NHEK. Since our solutions have a bubbling structure in the IR, they correspond to an IR phase of broken conformal symmetry, and their existence supports the possibility that all the pure states whose counting gives the Kerr black hole entropy correspond to horizonless bulk configurations.

\end{adjustwidth}

\thispagestyle{empty}
\newpage


\baselineskip=14pt
\parskip=3pt

\setcounter{tocdepth}{2}
\tableofcontents

\baselineskip=15pt
\parskip=3pt



\section{Introduction}

The construction and analysis of smooth solitonic geometries in supergravity have attracted an intense activity over the last few years driven by the fuzzball proposal \cite{Mathur:2005zp}. This proposal postulates that the semi-classical picture of black hole breaks down near its horizon and the black hole quantum state is a vector in a Hilbert space spanned by microstates approximated by smooth horizonless geometries that have the same mass, angular momentum and charges as the corresponding black hole. Most microstate geometries that have been constructed so far correspond to supersymmetric extremal black holes \cite{Bena:2004de,Berglund:2005vb,Bena:2007kg,Bena:2010gg,Bena:2015bea}. This domain of research brought many fruitful developments as the construction of the entropy enigma \cite{Denef:2007vg}, the calculation of index-jumps when crossing walls of marginal stability \cite{Sen:2007vb}, their application to AdS$_2$ holography \cite{Bena:2018bbd}, the quantization of the phase space of multicenter solution \cite{deBoer:2008zn} and many others. 

From this large amount of breakthroughs, it is natural to take this line of research further towards the description of microstates of real astrophysical black holes. Several steps were taken in the construction of non-extremal non-supersymmetric black hole microstates \cite{Bossard:2014ola,Bena:2015drs,Bena:2016dbw,Bossard:2017vii}. 

In the present work, we are interested in the extreme Kerr black hole. Such a black hole in four dimensions has an angular momentum $J$ which saturates the bound $J \leq G M^2$ and its near-horizon geometry contains a specific  warped AdS$_3$ factor (WAdS$_3$), which is a particular U(1) fiber over AdS$_2$. 

The Kerr/CFT correspondence has been first conjectured in \cite{Guica:2008mu} and relates the near-horizon geometry of extremal Kerr black hole (NHEK) to a chiral 2-dimensional conformal CFT whose the central charges are given by the angular momenta of the black hole. This conjecture correctly reproduces via Cardy's formula the Bekenstein-Hawking entropy of the black hole. Nevertheless, even if there are several possible candidates of dual ``CFT$_2$" as a dipole CFT \cite{ElShowk:2011cm} or as warped-CFT \cite{Detournay:2012pc}, the Kerr/CFT holographic dictionary is still poorly understood. Hence, it is very useful to have concrete examples, if not of the CFT, then of asymptotically NHEK geometries, which are bulk duals of pure CFT states.

For this purpose, it is crucial to have embeddings of NHEK in string theory as geometries arising from a system of D-branes. One of these is the six-dimensional uplift of the extremal non-supersymmetric D1-D5-P black hole in type IIB string theory on $T^4\times S^1$ \cite{Breckenridge:1996sn,Cvetic:1996xz,Cvetic:1998xh}. This theory contains the extremal non-supersymmetric Kerr-Newman black hole solution with one of its angular momenta set to be zero. Its near-horizon geometry is a squashed S$^3$ (SqS$^3$) over WAdS$_3$ which corresponds to a NHEK geometry, but the warp factor is constant, unlike for the NHEK solution in four dimensions \footnote{In 4d NHEK the warp factor depends on the polar angle.}.

In \cite{Bena:2012wc}, it has been shown that WAdS$_3\times$SqS$^3$ solutions, of which the NHEK spacetime is a particular example, can be obtained from AdS$_3\times$S$^3$  by a specific sequence of supergravity transformations known as $\mathcal{STU}$ transformations or generalized spectral flows \cite{Bena:2008wt}. This sets up the first cornerstone to build more solutions with a NHEK region in supergravity since they can be generated from more ``common" six-dimensional non-supersymmetric extremal solutions in type IIB string theory (see \cite{Bena:2015pua} for instance).

The main goal of this paper is to apply this technique to construct smooth bubbling geometries which are asymptotically WAdS$_3\times$SqS$^3$ or more particularly asymptotically NHEK. Our methodology is to start from a family of initial non-supersymmetric extremal solutions known as almost-BPS multicenter solutions \cite{Goldstein:2008fq,Bena:2009en}. As their BPS cousins, they are defined by a certain number of centers in a four-dimensional Taub-NUT space which carry magnetic and electric charges corresponding to branes wrapping cycles of the transverse space. The supersymmetry is broken in a subtle way by having opposite duality between the Taub-NUT space and the  fluxes \cite{Goldstein:2008fq}. Their conditions of existence being close to the BPS multicenter solutions, one can easily generate families of almost-BPS solutions using similar technique as for BPS solutions. Following the idea of \cite{Heidmann:2017cxt}, we will work with the family of almost-BPS solutions with three supertube centers in Taub-NUT. Each center preserves locally 16 supersymmetries. One can systematically construct such solutions and their parameter space is well-understood. Initially, these solutions are not regular in six dimensions since each species of supertube sources a different KKM dipole charge. However, as explained in \cite{Bena:2008wt}, the three generalized spectral flows transform each of the three supertube centers to a smooth Gibbons-Hawking center. We then expect that the spectrally-flowed solutions which are the ones containing the NHEK will be smooth.

If the generalized spectral flows map one BPS solution to another, they have a much richer structure for almost-BPS solutions as the spectrally flowed solutions do not belong to the almost-BPS class any more \cite{DallAgata:2010srl}. By deriving the effect of generalized spectral flows on our three-supertube solutions, we will show in section \ref{sec:Bubblinggeneralities} that one can indeed obtain systematically asymptotically WAdS$_3\times$SqS$^3$ solutions by just constraining the initial solutions to have both angular momenta to be zero in order to be asymptotic to the specific U(1) fiber over AdS$_2$ that gives the full AdS$_3$.

Having the same metric at infinity is not the only requirement to build either asymptotically WAdS$_3\times$SqS$^3$ or asymptotically NHEK solutions. The periodicities of the angles of the squashed 3-sphere and the angle of the warped AdS$_3$ must have a specific form (equation \eqref{eq1:NHEKperiodicities} for WAdS$_3$ with the specification \eqref{eq1:NHEKperiods} for NHEK). Imposing such periodicities to our solutions in the UV has a major impact on the smoothness of the geometry in the IR. Indeed, multicenter solutions have a $\mathbb{R}^{1,4}\times$S$^1$ local geometry around each center. Thus, depending on the periodicities of the angles, conical defects can occur at these locations. A tedious smoothness analysis needs to be performed to have a smooth discrete quotient of $\mathbb{R}^{1,4}\times$S$^1$ at each center.
\begin{figure}[t]
\begin{center}
\begin{tikzpicture}
 \draw (-7.5,4.2) rectangle (-2.5,5.5)  node[pos=.5](ABPS) {Almost-BPS solutions};
\draw (-5,-3.5) node {Supertube bubbles};
\draw (-5,3.2) node {AdS$_3\times$S$^3$};
\node[inner sep=-2pt] (russell) at (-5,0)
    {\includegraphics[height=0.4\textwidth]{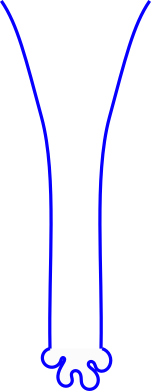}};
\node[inner sep=-2pt] (whitehead) at (5,0)
    {\includegraphics[height=0.4\textwidth]{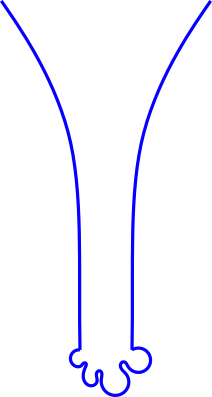}};
\draw[->,line width=0.5mm] (-2.8,0) -- (2.8,0)
    node[above,midway,fill=white] {3 generalized}    node[below,midway,fill=white] {spectral flows};
\draw (1.7,4.2) rectangle (8.5,5.5) node[pos=.5](NHEK) {Asymptotically NHEK solutions};
\draw (5,-3.5) node {Smooth bubbles};
\draw (5,3.2) node {NHEK};
\draw[->,line width=0.2mm] (-1.8,4.8) -- (1,4.8);
\end{tikzpicture}
\caption{Schematic description of the construction of a family of smooth asymptotically NHEK solutions from the family of almost-BPS solutions.}
\label{fig:Procedurepicture}
\end{center}
\end{figure}
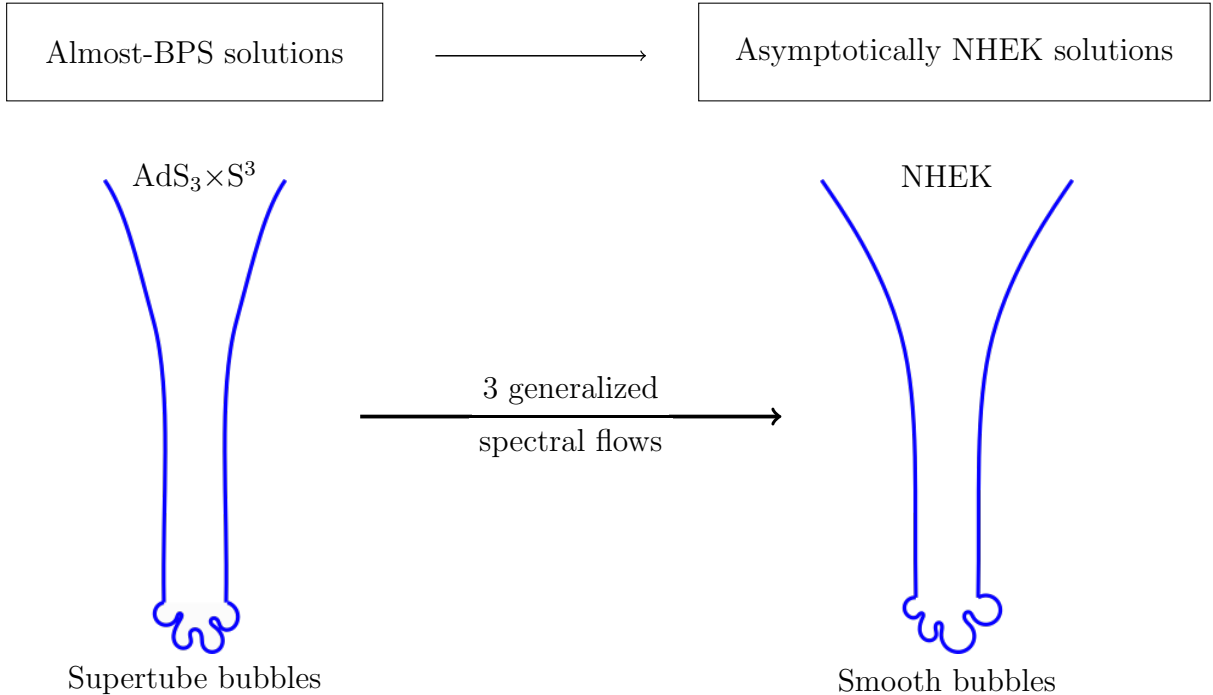

We sketch the overall idea about our construction of smooth bubbling asymptotically WAdS$_3\times$SqS$^3$ or asymptotically NHEK geometries in Figure \ref{fig:Procedurepicture}. The recipe we use to construct the solutions has the following steps:
\begin{itemize}[noitemsep,topsep=0pt]
\setlength{\itemsep}{0.2\baselineskip}
\item We start with a specific family of almost-BPS solutions. They have four centers, one is the center of Taub-NUT and the three others are two-charge supertube centers of different species. This choice of solution is just a matter of simplicity since a systematic construction is given in \cite{Heidmann:2017cxt} when the solutions are supersymmetric. However, nothing prevents from taking different almost-BPS configurations. We construct asymptotically AdS$_2\times$S$^1\times$S$^3$ solutions where the S$^1$ fibration over AdS$_2$ gives the full AdS$_3$. Moreover, we require their left and right angular momenta to be zero.
\item We perform three generalized spectral flows which transform the UV geometry to WAdS$_3\times$SqS$^3$ and preserve the bubble feature in the IR.
\item We identify the periodicities at infinity of WAdS$_3\times$SqS$^3$ or NHEK and impose the absence of conical singularities at the centers.
\end{itemize}

Once all these steps are performed, we obtain a family of smooth bubbling geometries, asymptotically WAdS$_3\times$SqS$^3$ or asymptotically NHEK in type IIB on T$^4$. From a Kerr/CFT perspective, these bubbling geometries are dual to CFT pure states of the chiral CFT$_2$ dual to WAdS$_3$ or NHEK.

Nearly extreme Kerr black holes have been observed in the sky (\cite{McClintock:2006xd} for instance). Thus, our construction can also lead to interesting astrophysical computations. One can compute the Kerr multipole moments of our solutions to see if there exist deviations from the Kerr black hole solution. This could give interesting observable imprints of the microstate structure of black holes on the gravitational wave emission after a collision between two black holes, which are expected to be measurable with LISA.

In section \ref{sec:5DD1D5PBH}, we review the six-dimensional uplift of the non-supersymmetric rotating D1-D5-P black holes in type IIB string theory on $T^4$. We discuss their near-horizon geometry and the specific transformations which deform AdS$_3$ to WAdS$_3$. In section \ref{sec:Bubblinggeneralities}, we review the almost-BPS solutions in type IIB, the effect of generalized spectral flows on multicenter solutions and we discuss in detail the systematic construction of our starting family of almost-BPS solutions. In section \ref{sec:bubblingWAdS} we apply the procedure discussed above to construct asymptotically WAdS$_3$ smooth bubbling solutions and in section \ref{sec:bubblingNHEK} we construct similarly asymptotically NHEK smooth bubbling solutions. In both sections, we give explicit examples of solutions.
\newpage

\section{5D extremal rotating black holes in type IIB}
\label{sec:NHEKgeneralities}

In this section we briefly review the description of non-supersymmetric three-charge over-rotating Cvetic-Youm black holes which arise in the low-energy limit of type IIB string theory compactified on $T^4\times S^1$ \cite{Breckenridge:1996sn,Cvetic:1996xz,Cvetic:1998xh}. We describe their near-horizon or NHEK limit and the more general family of warped-AdS$_3$ geometries (WAdS$_3$) to which those NHEK solutions belong to \cite{Dias:2007nj, Guica:2010ej, Song:2011ii,ElShowk:2011cm, Bena:2012wc}. We also discuss the supergravity transformations which deform an AdS$_3$ geometry to a WAdS$_3$ geometry \cite{Bena:2012wc}.

\subsection{The non-supersymmetric extremal D1-D5-P black holes}
\label{sec:5DD1D5PBH}

We work in the context of type IIB string theory on a $T^4\times S^1$. We assume that the torus is much smaller than the one-sphere. As a consequence, the five-dimensional black hole solutions can be seen as six-dimensional black string solutions. In this paper, we consider a four-parameter family of non-supersymmetric extremal spinning black holes characterized by a mass $M$, two $SU(2)_L$ and $SU(2)_R$ angular momenta $J_L$ and $J_R$ and three charges $Q_I$ with $I=1,5,p$ as follows
\begin{equation}
\begin{split}
M ~&=~ 2\, a^2 \left(c_1^2 + s_1^2 + c_5^2 + s_5^2 + c_p^2 + s_p^2  \right), \\
J_R ~&=~ 0,\\
J_L ~&=~ 4 \,a^3 \left(c_1 c_5 c_p + s_1 s_5 s_p  \right), \\
Q_I ~&=~4 \, a^2 \,s_I c_I , \qquad I=1,5,p,
\end{split}
\label{eq1:D1D5PchargesmassAM}
\end{equation}
where $s_I = \sinh \delta_I$ and $c_I = \cosh \delta_i$. The Bekenstein-Hawking entropy and the left and right temperatures are
\begin{equation}
\begin{split}
S_{BH} ~&=~2\pi \sqrt{J_L^2 - Q_1 Q_5 Q_p} ~=~8 \pi \,a^3 \left(c_1 c_5 c_p - s_1 s_5 s_p  \right), \\
T_L ~&=~ 0, \\
T_R ~&=~ \frac{1}{\pi} \sqrt{1 - \frac{Q_1 Q_5 Q_p}{J_L^2}}. \\
\end{split}
\end{equation}
The metric of the six-dimensional black string is \cite{Cvetic:1998xh,Dias:2007nj}

\begin{alignat}{1}
ds_6^2 ~=~ &-\left(1-\frac{4a^2 c_p^2}{\hat{r}^2+a^2} \right) \frac{d\hat{t}^2}{\sqrt{H_1 H_5}} +\left(1+\frac{4a^2 s_p^2}{\hat{r}^2+a^2} \right) \frac{d\hat{y}^2}{\sqrt{H_1 H_5}} \nonumber\\
&+\frac{8 a^2 s_p c_p}{(\hat{r}^2+a^2)\sqrt{H_1 H_5}} \,d\hat{t}\,d\hat{y} + (\hat{r}^2+a^2)\sqrt{H_1 H_5} \left(\frac{\hat{r}^2}{(\hat{r}^2-a^2)^2}\,d\hat{r}^2+d\hat{\theta}^2 \right. \label{eq1:D1D5p6Dmetric} \\
&\left. +\cos^2\hat{\theta} \,d\hat{\psi}^2 + \sin^2 \hat{\theta} \,d\hat{\phi}^2 \right) + \frac{4 a^4}{(\hat{r}^2+a^2)\sqrt{H_1 H_5}} \left(\cos^2\hat{\theta} \,d\hat{\psi} + \sin^2 \hat{\theta} \,d\hat{\phi} \right)^2 \nonumber\\
&- \frac{8 a^3}{(\hat{r}^2+a^2)\sqrt{H_1 H_5}}\left((c_1c_5c_p+s_1s_5s_p)dt +(s_1s_5c_p +c_1c_5s_p)d\hat{y} \right)\left(\cos^2\hat{\theta} \,d\hat{\psi} + \sin^2 \hat{\theta} \,d\hat{\phi} \right) \nonumber
\end{alignat}
where
\begin{equation}
H_i ~=~ 1 + \frac{4a^2 s_i^2}{\hat{r}^2+a^2},\qquad i=1,5. 
\end{equation}
The coordinate $\hat{r}$ is the radial coordinate of the four-dimensional base space defined by $\hat{r}$ and the three angles $\hat{\theta}$, $\hat{\phi}$, $\hat{\psi}$ and $\hat{y}$ is the KK direction. The periodicities of the angles $\hat{y}$, $\hat{\psi}$ and $\hat{\phi}$ are\footnote{We have chosen the unit $R_{\hat{y}}=1$.}
\begin{equation}
(\hat{y},\hat{\psi},\hat{\phi}) =
\left\{
\arraycolsep=1.0pt\def\arraystretch{1.6}
\begin{array}{rl}
(\hat{y},\hat{\psi},\hat{\phi}) \,\,+\,\, &2\pi \,(1 , 0  ,0)\\
(\hat{y},\hat{\psi},\hat{\phi}) \,\,+\,\, & 2\pi \,(0, 1  ,0)\\
(\hat{y},\hat{\psi},\hat{\phi}) \,\,+\,\, &2\pi \,(0 , 0  ,1)
\end{array}
\right. ,
\label{eq1:D1D5PBHperiodicities}
\end{equation}
The geometry is asymptotically $\mathbb{R}^{1,4}\times S^1$ and has an event horizon at $\hat{r}_h = a$.
\subsection{The near-horizon extremal Kerr geometry}

The near-horizon geometry of the six-dimensional uplift of the five-dimensional black hole solutions given in \eqref{eq1:D1D5p6Dmetric} has been shown to be a \emph{near-horizon extremal Kerr} geometry (NHEK) \cite{Dias:2007nj,Bredberg:2009pv,Guica:2010ej,ElShowk:2011cm,Bena:2012wc}. It is a squashed $S^3$ fibered over warped AdS$_3$ with specific angle periodicities. 

\noindent
The near-horizon limit is obtained by changing the coordinates as follows
\begin{alignat}{3}
t ~&=~ \frac{4 \pi \, \epsilon}{S}\, \hat{t}, \qquad &&r~=~ \frac{\hat{r}^2-a^2}{\epsilon} ,\qquad &&y ~=~ \frac{S}{\pi Q_1 Q_5}\left(\hat{y}-V_H \hat{t}\right),\nonumber\\
\psi ~&=~ \hat{\psi} + \hat{\phi} - \frac{8 \pi \,a^2}{S} \, \hat{t}-\frac{4 J_L}{Q_1 Q_5} \left(\hat{y}-V_H \hat{t}\right),\qquad &&\phi ~=~   \hat{\psi} - \hat{\phi} ,\qquad &&\theta ~=~ 2 \,\hat{\theta},
\end{alignat}
where $V_H$ is the linear velocity $V_H=-\frac{8\pi\,a^3(c_1c_5s_p-s_1s_5c_p)}{S}$. Thus, the periodicities of $y$, $\psi$ and $\phi$ are given by the following identifications
\begin{equation}
(y,\psi,\phi) =
\left\{
\arraycolsep=1.6pt\def\arraystretch{1.6}
\begin{array}{rl}
(y,\psi,\phi)  \,\,+\,\, &2\pi \,(T_y , -T_\psi  ,0)\\
(y,\psi,\phi)  \,\,+\,\, & 2\pi \,(0, 2  ,0)\\
(y,\psi,\phi)  \,\,+\,\, &2\pi \,(0 , 1  ,1)
\end{array}
\right. ,
\label{eq1:NHEKperiodicities}
\end{equation}
with 
\begin{equation}
T_y ~\equiv~ \frac{S}{\pi Q_1 Q_5}\, , \qquad T_\psi ~\equiv~ \frac{2 J_L}{Q_1 Q_5}\,.
\label{eq1:NHEKperiods}
\end{equation}
By taking the limit $\epsilon \rightarrow 0$, the near-horizon metric in terms of the above coordinates gives rise to a T$^2$ fibration over AdS$_2\times$ S$^2$
\begin{equation}
\begin{split}
ds_{\text{NHEK}}^2 ~=~ \frac{\kappa^2}{4} &\left[-r^2 dt^2 + \frac{dr^2}{r^2} \,+\, \gamma \,(dy +r dt)^2 + \gamma \, ( d\psi + \cos\theta  d\phi )^2   \right. \\
 &\left.   \:+\: 2 \alpha \, (dy +r dt) ( d\psi + \cos\theta \, d\phi ) \,+\, d\theta^2 + \sin^2 \theta \,d\phi^2\right. \bigg].
\end{split}
\label{eq1:NHEKWAdSmetric}
\end{equation}
where the warp constant factors $\gamma$ and $\alpha$ and the length $\kappa$ are given by
\begin{equation}
\begin{split}
\gamma &\,=\, 1 + \frac{1}{(c_1^2 +s_1^2)(c_5^2 + s_5^2)} \,,\qquad \alpha \,=\, \frac{1}{c_1^2 +s_1^2} + \frac{1}{c_5^2 +s_5^2}\,,\\
\kappa^2 &\,=\, 2 a^2 \sqrt{(c_1^2 +s_1^2)(c_5^2 + s_5^2)}\,.
\end{split}
\label{eq1:WAdSfactors}
\end{equation}

This background belongs to the family of deformations of AdS$_3\times$S$^3$ into squashed S$^3$ (SqS$^3$) over warped AdS$_3$ (WAdS$_3$) \cite{Anninos:2008fx,Orlando:2010ay,Song:2011sr}. However, for generic WAdS$_3\times$SqS$^3$ backgrounds, the periods $T_y$ and $T_\psi$ \eqref{eq1:NHEKperiodicities} are arbitrary. 

\noindent
One can rewrite the solutions in terms of the SU(2)$_L$-invariant one-forms on S$^3$ 
\begin{equation}
\begin{split}
&\sigma_1 \,=\, \cos \psi \, d\theta + \sin\theta \sin \psi \,d\phi\, , \qquad \sigma_2 \,=\, -\sin \psi \, d\theta + \sin\theta \cos \psi \,d\phi\, ,  \\
&\sigma_3 \,=\, d\psi+ \cos\theta\,d\phi\, , 
\end{split}
\end{equation}
and the SL(2,$\mathbb{R}$)$_L$-invariant one forms on AdS$_3$ 
\begin{equation}
w_+ \,=\, -e^{-y} \left(\frac{dr}{r} +r dt \right)\, , \qquad w_- \,=\, e^{y} \left(\frac{dr}{r} - r dt \right)\, , \qquad w_3 \,=\, dy +r dt,
\end{equation}
to make the WAdS$_3\times$SqS$^3$ geometry manifests
\begin{equation}
\begin{split}
ds_{NHEK}^2 ~=~ \frac{\kappa^2}{4} \,\left( -w_+ w_- + \gamma \, w_3^2 +\sigma_1^2 +\sigma_2^2 + \gamma \, \sigma_3^2 +  2\alpha \, w_3 \sigma_3\right).
\end{split}
\end{equation}
In the present paper, we want to build asymptotically WAdS$_3\times$SqS$^3$ and more particularly asymptotically NHEK bubbling geometries. For that purpose, it is interesting to review the supergravity transformations described in \cite{Bena:2012wc} which deform AdS$_3\times$S$^3 \rightarrow$ WAdS$_3\times$SqS$^3$ since building asymptotically AdS$_3\times$S$^3$ geometries is well-controlled and well-known.

\subsection{From AdS$_3\times$S$^3$ to  WAdS$_3\times$SqS$^3$ }
\label{sec:adstowads}

The sequence of supergravity transformations from an AdS$_3\times$S$^3$ spacetime to a WAdS$_3\times$SqS$^3$ has been exhaustively detailed in \cite{Bena:2012wc}. We just give a brief summary in this section. The transformations can be seen as a series of $\mathcal{STU}$ transformations or equivalently as a sequence of three generalized spectral flows:
\begin{itemize}
\item[-] A $\mathcal{T}$ Kähler transformation or the first generalized spectral flow.

 A $\mathcal{T}$ Kähler transformation consists of a T-duality along $y$ followed by a shift of $\varphi\rightarrow \varphi + \gamma y$ where $\varphi$ can be either $\psi$ or $\phi$ and a T-duality back along $y$. We usually denote such a transformation as ``TsT". The first generalized spectral flow differs from $\mathcal{T}$ transformation by a S-duality at the beginning and at the end (STsTS) and it induces the same transformation of the geometry. 

\item[-] A $\mathcal{S}$ transformation or the second generalized spectral flow.

Using the previous notation, the second generalized spectral flow can be denoted as a T$^4$STsTST$^4$ transformation where the ``T$^4$" part refers to four T-dualities on the transverse 4-torus. A $\mathcal{S}$ transformation corresponds to the same transformation with an S-duality at the beginning and at the end.

\item[-] A $\mathcal{U}$ transformation or the third generalized spectral flow.

The transformations correspond to a volume-preserving change of coordinate which simply corresponds to a shift ``s" of $\varphi$.
\end{itemize}

The two possible choices of $\varphi$ differ significantly. If the AdS$_3\times$S$^3$ background has a dual three-form field strength, the $\mathcal{STU}$ transformations associated to this background (SL(2,$\mathbb{R}$)$_L\times$SU(2)$_L$ invariant) or the generalized spectral flows along $\psi$ will preserve supersymmetry and the transformed geometry will remain AdS$_3\times$S$^3$. Reversely, if the three-form field strength is anti-self-dual, the $\mathcal{STU}$ transformations associated to this background (SL(2,$\mathbb{R}$)$_L\times$SU(2)$_R$ invariant) or the generalized spectral flows along $\phi$ will break supersymmetry and will transform the geometry to WAdS$_3\times$SqS$^3$ (see section 2.2 of \cite{Bena:2012wc} for more precision).


\section{Almost-BPS bubbling geometries in type IIB}
\label{sec:Bubblinggeneralities}

In the previous section, we have described the UV geometry we want to build. We have detailed how it can be obtained from an AdS$_3\times$S$^3$ UV geometry by spectral flow transformations. In the current section, we discuss our choice of non-supersymmetric extremal bubbling geometries. We work with a family of ``almost-BPS" multicenter solutions in type IIB string theory on a T$^4\times$S$^1$ \cite{Goldstein:2008fq,Bena:2009en,Bena:2009ev,DallAgata:2010srl,Vasilakis:2011ki}. We will be brief in the review of their general construction in section \ref{sec:generalitiesABPSmulticenter} as these solutions are discussed in great detail in \cite{Bena:2009en}. In section  \ref{sec:asymptotics} we explain the construction of asymptotically AdS$_3\times$S$^3$ almost-BPS solutions. Then, we review the effect of generalized spectral flows on these geometries in section \ref{sec:asymptotics} \cite{DallAgata:2010srl}. Finally, we focus on the particular family of three almost-BPS two-charge supertubes in $\mathbb{R}^4$ in section \ref{sec:3SinTNgeneralities}. 

\subsection{Multicenter solutions in Taub-NUT}
\label{sec:generalitiesABPSmulticenter}

The extremal almost-BPS solutions are constructed with the same ansatz of type IIB metric and matter fields as the BPS solutions:
\begin{alignat}{1}
ds^2_{\text{ABPS}} &~=~- \frac{1}{Z_p \sqrt{Z_1 Z_5}} (dt+k)^2 +\sqrt{Z_1 Z_5} ds_4^2 +  \frac{Z_p}{ \sqrt{Z_1 Z_5}} (A_p + dy)^2 + \sqrt{\frac{Z_1}{Z_5}}ds_{T_4}^2, \nonumber \\
e^{2\Phi} &~=~ \frac{Z_1}{Z_5}, \label{eq2:ABPSansatz}\\
B^{(2)} &~=~ 0, \nonumber \\
A_I  &~=~ -\frac{dt+k}{Z_I} + a_I, \qquad I=1,5,p,\nonumber
\end{alignat}
where $Z_I$ are the warp factors with $I=1,5,p$ encoding respectively the electric D1, D5 and P charge, $a_I$ are the magnetic one-forms, $k$ the angular momentum one-form, $A_I$ are the electromagnetic gauge fields and $ds_4$ is a hyper-Kähler four-dimensional metric which is chosen to have a Gibbons-Hawking form \cite{Gibbons:2013tqa, Niehoff:2016gbi}
\begin{equation}
\label{eq2:GHmetric}
ds_4^2 ~=~  V^{-1} \left( d\psi+ A\right)^2 + V \left( dr^2 + r^2( d\theta^2 + \sin^2 \theta d\phi^2 )\right) \, , \qquad \star_{3} dA=d V \, .
\end{equation}
The Hodge star $\star_3$ is with respect to the three-dimensional base space, the one-form $A$ is a Kaluza-Klein gauge field and $V$ is the Taub-NUT potential:
\begin{equation}
V ~=~ h_\infty + \frac{q}{r} \,~\Longrightarrow~ A ~=~ q \cos\theta \, d\phi.
\label{eq2:GHfunction}
\end{equation}
The RR three-form flux is given by
\begin{equation}
F^{(3)} ~=~dA_1 \wedge (A_p + dy) - \left(\frac{{Z_5}^5}{{Z_1}^3{Z_p}^2} \right)^{1/4} \star_5 dA_5 \,.
\label{eq2:RRthreeformFlux}
\end{equation}
\noindent
The almost-BPS equations of motion are
\begin{equation}
\begin{split}
& d a_I ~=~ -\star_4 \,  d a_I, \\
& d\star_4d Z_I ~=~ \frac{|\epsilon_{IJK}|}{2} \, d a_J \wedge d a_K, \\
& dk - \star_4 dk ~=~Z_I d a_I,
\end{split}
\end{equation}
where $\star_4$ is the Hodge star with respect to the Gibbons-Hawking space and $\epsilon_{IJK}$ is the Levi-Civita tensor with $\epsilon_{15p}=1$. An almost-BPS background breaks supersymmetry by reversing the duality of the magnetic dipole field strengths $da_I$ and of the angular momentum one-form $k$ (anti-self-dual) relative to the duality of the curvature of the Gibbons-Hawking space (self-dual). We expand $a_I$ and $k$ along the $\psi$-fiber of the Gibbons-Hawking space
\begin{equation}
a_I ~=~K_I (d\psi+A) +w_I\, \qquad k~=~ \mu (d\psi +A) + \omega.
\end{equation}
The equations of motion become equations on the three-dimensional base space
\begin{equation}
\begin{split}
& d\star_3d Z_I ~=~  \frac{|\epsilon_{IJK}|}{2} \, V \,d\star_3d (K_J K_K ), \\
& \star_3d w_I ~=~ V \,dK_I - K_I \,dV, \\
& \star_3d \omega ~=~ V Z_I dK_I- d(\mu V),\\
& d\star_3 d (\mu V) ~=~ -d(V Z_I) \star_3 dK_I 
\label{eq2:EOM}
\end{split}
\end{equation}
The solutions are determined by eight harmonic functions $\{V,K_1,K_5,K_p, L^1, L^5, L^p,M\}$ where the functions $L_I$ source the warp factors $Z_I$ and $M$ sources $\mu$. Each harmonic function is sourced by $n+1$ centers on the three-dimensional base space. In the present paper, we are interested in axisymmetric configurations where the centers are denoted by a coordinate $a_i$ on the $z$ axis in $\mathbb{R}^3$, $i=0...n$ with $a_0=0$,. The harmonic functions carry a charge at each center. We use the following notation:
\begin{equation}
K^I=k^I_\infty+\sum_{i=0}^n \frac{k^I_i}{r_i} \, , \hspace{0.3cm}
L_I=l^I_\infty+\sum_{i=0}^n \frac{Q^{(I)}_i}{r_i} \, , \hspace{0.3cm}
M=m_\infty+\sum_{i=0}^n \frac{m_i}{r_i} \, ,
\label{eq2:harmfunc}
\end{equation}
where $r_i$ is the three-dimensional distance to the $i^{th}$ center $r_i = \sqrt{r^2 +a_i^2 - 2 r a_i \cos\theta}$. From those expressions and the equations \eqref{eq2:EOM}, one can derive the general form of the warp factors \cite{Bena:2009en}
\begin{equation}
Z_I ~=~ L_I +\frac{|\epsilon_{IJK}|}{2} \sum_{j,k} \left(h_\infty + \frac{q r}{a_j a_k}\right)\frac{k^J_j k^K_k}{r_j r_k},
\label{eq2:Zgeneralform}
\end{equation}
and the generic expression of the angular momentum one-form $k$ given by $\omega$ and $\mu$ can be found in \cite{Bena:2009en}.

In anticipation of the computation of generalized spectral flows in section \ref{sec:ABPSspectralflows}, we define four magnetic and electric one-forms, $v_I$ and $v_0$, determined by the following equations
\begin{equation}
\begin{split}
\star_3 dv_I &~\equiv~ -dZ_I + \frac{|\epsilon_{IJK}|}{2} \left(V d(K_J K_K) - K_J K_K dV \right), \\
\star_3 dv_0 &~\equiv~ K_I dZ_I - Z_I dK_I +K_1 K_5 K_p dV -  V d(K_1 K_5 K_p).
\end{split}
\label{eq3:v0vIequations}
\end{equation}

These solutions do not necessarily correspond to physical geometries. For that purpose, regularity conditions have to be satisfied \cite{Bena:2009en}:
\begin{itemize}
\item[-] Since the angular momentum one-form $\omega$ is proportional to $d\phi$, $\omega$ must vanish on the z-axis where $\phi$ degenerates to avoid Dirac-Misner string singularities. This imposes $n+1$ bubble equations on the distances between the centers (see \cite{Bena:2009en} for the generic equations).
\item[-] The absence of closed timelike curves requires the positivity of the quartic invariant:
\begin{equation}
\mathcal{I}_4 ~\equiv~ Z_1 Z_5 Z_pV - \mu^2 V^2 ~>~ 0.
\label{eq2:I4}
\end{equation}
\end{itemize}

Those regularity conditions constrain significantly the parameter space of the solutions. Once they are satisfied, we have a family of extremal non-supersymmetric solutions in six dimensions which cap off in the IR. The solutions are regular everywhere but they might have singularities at the centers. These corresponding regularity conditions depend on the nature of the centers and will be discussed for supertube centers in section \ref{sec:regularity3S}.


\subsection{Asymptotics of multicenter solutions}
\label{sec:asymptotics}

In section \ref{sec:adstowads} we have described a procedure to go from an AdS$_3\times$S$^3$ UV region to WAdS$_3\times$SqS$^3$. We discuss now the asymptotics of a bubbling almost-BPS solution. We derive the conditions to be asymptotic to the specific S$^1$ fibration over AdS$_2\times$S$^3$ that gives the full AdS$_3\times$S$^3$:
\begin{equation}
ds_{\infty}^2 ~\propto~ -r^2 dt^2 + \frac{dr^2}{r^2} \,+\,(dy +r dt)^2 + d\Omega^2_3.
\label{eq2:targetAdSmetric}
\end{equation}
The asymptotics of a multicenter solution is given by the large-distance behavior of the warp factors $Z_I$, the Gibbons-Hawking function $V$ and the angular momentum one-form $k$. We already assume that the constant term in $V$ is zero which is a straightforward necessary condition to have an AdS factor at infinity. The series expansion of $Z_I$, $V$ and $k$ involves the constant terms $l^I_\infty$ and $m_\infty$, the D1, D5 and P charges and the left and right angular momenta of the solution which we denote as $q_1$, $q_5$, $q_p$, $j_L$ and $j_R$:
\begin{equation}
\begin{split}
Z_1 &~\underset{r \rightarrow\infty}{\sim}~ \l^1_\infty \,+\, \frac{q_1}{r}, \qquad \quad Z_5 ~\underset{r \rightarrow\infty}{\sim}~ \l^5_\infty \,+\, \frac{q_5}{r}, \qquad \quad  Z_p ~\underset{r \rightarrow\infty}{\sim}~ \l^p_\infty \,+\, \frac{q_p}{r}, \\
k &~\underset{r \rightarrow\infty}{\sim}~ \frac{j_R + j_L \cos \theta}{r}\, d\psi \:+\: q \,\frac{j_L +j_R \cos\theta}{r}\, d\phi, \qquad  V ~\underset{r \rightarrow\infty}{\sim}~ \frac{q}{r}.
\end{split}
\label{eq3:asymptoticbehaviorWFV}
\end{equation}
From the metric \eqref{eq2:ABPSansatz}, we see that an asymptotic behavior as \eqref{eq2:targetAdSmetric} can be achieved by imposing that \emph{all the constant terms and the left and right angular momenta are strictly zero}\footnote{Having no angular momentum essentially means that $\mu$ and $\omega$ decay as $r^{-2}$}.

The requirement $h_\infty=0$ makes the Taub-NUT space to be a trivial $\mathbb{R}^4$. The supersymmetry breaking obtained by an opposite direction between the base space and the gauge fields does not hold anymore and the solution can be mapped to a BPS solution by interchanging $\phi \leftrightarrow \psi$ \cite{Bena:2012wc,Goldstein:2008fq,Bena:2009ev}. However, as explained in section \ref{sec:adstowads}, one can still count on the generalized spectral flows to break supersymmetry.

Furthermore, in our language, having no constant terms in the harmonic functions makes the solution to be asymptotically AdS$_2\times$S$^1$ rather than AdS$_3$ \cite{Bena:2018bbd}. This is just a matter of convention since one can consider the AdS$_3\times$S$^3$ metric \eqref{eq2:targetAdSmetric} as a U(1) fiber on an AdS$_2\times$S$^3$.

Once the conditions on the constant terms and the angular momenta are satisfied, one can perform a sequence of three generalized spectral flows. They will transform the IR geometry by keeping it bubbling and by transforming the supertube center to smooth center \cite{Bena:2008wt}. They will transform the UV geometry from \eqref{eq2:targetAdSmetric} to a WAdS$_3\times$SqS$^3$ geometry \eqref{eq1:NHEKWAdSmetric}.


\subsection{Almost-BPS generalized spectral flows}
\label{sec:ABPSspectralflows}

The three generalized spectral flows detailed in \ref{sec:adstowads} can be translated in the formalism of multicenter solutions as transformations of the NSNS and RR fields. As explained in the previous section, our choice of constant terms makes our solutions to be BPS by interchanging $\phi \leftrightarrow \psi$. Consequently, the generalized spectral flows corresponding to a shift $\phi \rightarrow \phi + \gamma y$ produce the usual BPS generalized spectral flows which consist in interchanging linearly the harmonic functions and which will preserve the supersymmetry \cite{Gauntlett:2004qy,Bena:2005ni,Bates:2003vx}. However, the generalized spectral flows corresponding to a shift $\psi \rightarrow \psi + \gamma y$ produce the expected supersymmetry breaking and will allow to go from the almost-BPS class of solutions to different non-supersymmetric classes \cite{Bena:2009fi,DallAgata:2010srl} as the family of asymptotically WAdS$_3$ solutions. We now review the transformation rules of generalized spectral flows on the NSNS and RR fields \cite{DallAgata:2010srl,Bena:2012wc}. Let us define the three constant shifts $\gamma_1$, $\gamma_5$ and $\gamma_p$ of the three types of spectral flows and the following new functions
\begin{equation}
T_I ~\equiv~ 1 + \gamma_I K_I \,, \qquad N_I ~=~ \frac{|\epsilon_{IJK}|}{2} \gamma_I^2 Z_J Z_K + VT_I^2 Z_I - 2\gamma_I V T_I \mu , \qquad I=1,5,p,
\end{equation}
and we define the usual short-hand notations $K^3= K_1 K_5 K_p$, $T^3= T_y T_5 T_p$, $N^3= N_1 N_5 N_p$, $\gamma^3= \gamma_1 \gamma_5 \gamma_p$ and $Z^3= Z_1 Z_5 Z_p$.
The spectrally flowed 6-dimensional metric and the matter gauge fields are given by \cite{DallAgata:2010srl}
\begin{alignat}{2}
d\widetilde{s}^2_{6d} &~=~ &&- \frac{1}{\widetilde{Z}_p \sqrt{\widetilde{Z}_1 \widetilde{Z}_5}} \left(dt+\widetilde{\mu}( d\psi+ \widetilde{A})+\omega\right) ^2+  \frac{\widetilde{Z}_p}{ \sqrt{\widetilde{Z}_1 \widetilde{Z}_5}} (\widetilde{A}_p + dy)^2\nonumber\\
& && +\sqrt{\widetilde{Z}_1 \widetilde{Z}_5}\left( \widetilde{V}^{-1} \left( d\psi+ \widetilde{A}\right)^2 + \widetilde{V} \, ds(\mathbb{R}^3)^2\right) , \label{eq3:SFmetric} \\
\widetilde{A}_I &~=~&& - \frac{dt +\omega}{\widetilde{W}_I} + \widetilde{P}_I ( d\psi+ \widetilde{A}) + \widetilde{w}_I,\nonumber
\end{alignat}
where 
\begin{alignat}{2}
\widetilde{V} &~=~ &&\left[ T^6 V^2 \,+\, 8 \gamma^3 T^3 V \mu \,-\, T^3 V \left(|\epsilon_{IJK}| \gamma_J \gamma_K T_I Z_I \right) \right. \nonumber \\
& && \left. \,+\, \frac{|\epsilon_{IJK}|}{2}\gamma_J^2 \gamma_K^2 T_I^2 Z_I^2 \,-\, |\epsilon_{IJK}| \gamma_I^2 \gamma_J \gamma_K T_J Z_J T_K Z_K\right]^{1/2}\, , \nonumber\\
\widetilde{A} &~=~ && A \,-\, \gamma_I w_I \,-\, \frac{|\epsilon_{IJK}|}{2} \gamma_J \gamma_K v_I \,+\, \gamma^3 v_0\, , \nonumber \\
\widetilde{Z}_I &~=~ && \frac{N_I}{\widetilde{V}} \, , \label{eq3:SFrules} \\
\widetilde{\mu} &~=~ && \widetilde{V}^{-2} \left(T^3 V^2 \mu - \gamma^3 Z^3 + \frac{ |\epsilon_{IJK}|}{2} \gamma_J \gamma_K Z_I T_I V \mu - \frac{ |\epsilon_{IJK}|}{2}  \gamma_I V T_J Z_J T_K Z_K \right)\,, \nonumber \\
\widetilde{W}_I &~=~ && \frac{N_I}{T^3 V +  \frac{ |\epsilon_{IJK}|}{2} \gamma_J \gamma_K T_I Z_I - |\epsilon_{IJK}| \gamma_I \gamma_J T_K Z_K} \,, \nonumber \\
\widetilde{P}_I &~=~ && \frac{V Z_I T_I K_I +  \frac{ |\epsilon_{IJK}|}{2} \gamma_I Z_J Z_K - (2 T_I-1)V \mu}{N_I}, \nonumber \\
\widetilde{w}_I &~=~ && w_I + |\epsilon_{IJK}| \gamma_J v_K -\frac{|\epsilon_{IJK}|}{2} \gamma_J \gamma_K v_0\,. \nonumber
\end{alignat}

Generalized spectral flows produce a non-trivial modification of the functions. However, the spectrally flowed extremal solutions still satisfy the regularity conditions. Indeed, the one-form $\omega$ is unchanged which guarantees the absence of Dirac-Misner string at $\theta=0,\pi$. Moreover, the quartic invariant is preserved under spectral flows $\tilde{\mathcal{I}}_4 = \mathcal{I}_4$. \emph{Hence, a regular almost-BPS multicenter solution is transformed by a generic spectral flows to a regular extremal non-supersymmetric solution}.

\noindent
Furthermore, if the initial almost-BPS solution has curvature singularities at the centers which happens for supertube centers of type $I=1,5$, the corresponding generalized spectral flows will transform the singular local geometries to quotients of $\mathbb{R}^{4}\times$S$^1$ \cite{Bena:2008wt}.

\noindent
Nevertheless, conical singularities related to the angle periodicities can still occur at these locations. Indeed, the NHEK angle periodicities or the WAdS$_3$ angle periodicities \eqref{eq1:NHEKperiodicities} imposed in the UV can spoil the periodicities at the centers where the three-sphere shrinks and conical singularities can emerge. Those are the only regularity conditions we need to worry about after spectral flows.

We now have all the basic ingredients to construct an extremal non-supersymmetric geometry which caps off smoothly in the IR and is asymptotically NHEK or WAdS$_3$. Working with the most generic almost-BPS multicenter solutions can lead to very complicated regularity conditions that are hard to analyze. That is why, we will focus our work on the family of four-center solutions of three two-charge supertubes in $\mathbb{R}^4$. As explain in \cite{Heidmann:2017cxt}, a systematic construction can be performed for BPS configurations. We extend the construction in the almost-BPS context in the next section.


\subsection{The family of almost-BPS three-supertube solutions in $\mathbb{R}^4$}
\label{sec:3SinTNgeneralities}

We consider the family of almost-BPS solutions with three two-charge supertubes and a $\mathbb{R}^4$ base space. For the BPS solutions, a systematic construction of the family have been performed \cite{Heidmann:2017cxt}. As explained in section \ref{sec:asymptotics}, almost-BPS solutions in $\mathbb{R}^4$ do not literally break supersymmetry as they can be mapped to BPS solutions \cite{Goldstein:2008fq,Bena:2009ev}. Those solutions are BPS, but can be constructed either as BPS or as almost-BPS solutions. Since we need the second for the spectral flows to break supersymmetry, we will do it here. 

\noindent
Thus, the extension of the construction is just a matter of rewriting carefully in the context of almost-BPS solutions. In this section, we apply the general results obtained in section \ref{sec:generalitiesABPSmulticenter} to our specific family of solutions. We will consider an axisymmetric supertube configuration. We first detail how the NSNS and RR fields are sourced by such a configuration. We then derive the regularity conditions and show that they can be systematically satisfied.

\subsubsection{The solution}

A type ``$I$" supertube, with $I=1,5,p$, has a singular magnetic source in $K^I$, two singular electric sources in $Z_J$ and $Z_K$ with $I \neq J \neq K$ and one angular-momentum charge in $M$ \cite{Mateos:2001qs}. The six-dimensional metric and the matter fields are still given by \eqref{eq2:ABPSansatz}. We assume that the $\mathbb{R}^4$ center is at the origin of the space and that a supertube of type 1 is at a second center with coordinate $a_1$ on the z-axis, a supertube of type 5 is at a third center with coordinate $a_5$ and a supertube of type p is at a fourth center with coordinate $a_p$. We consider that $a_I>0$. We denote by $r_I$ the three-dimensional distance from the $I^{th}$ center $r_I \equiv \sqrt{r^2 +a_I^2 - 2 r a_I \cos\theta}$. We use the following notation for the eight harmonic functions\footnote{We remind that all the constant terms in $V$ and $L^I$ have been set to zero to have an asymptotically AdS$_2\times$S$^1\times$S$^3$ solution.} 
\begin{alignat}{3}
V &~=~ \frac{q}{r} \, , \qquad && M &&~=~ m_{\infty}+\frac{m_0}{r} +\frac{m_1}{r_1}+\frac{m_5}{r_5} +\frac{m_p}{r_p} \,, \nonumber\\
K^{1} &~=~ k^1_\infty + \frac{a_1 \, \kappa_1}{q \, r_{1}} \, , \qquad &&L_{1} &&~=~ \frac{Q^{(1)}_5}{r_5} + \frac{Q^{(1)}_p}{r_p} \, , \label{eq3:initialsupsol}\\
K^{5} &~=~ k^5_\infty + \frac{a_5 \, \kappa_5}{q \, r_{5}}  \, , \qquad &&L_{5} &&~=~  \frac{Q^{(5)}_1}{r_1} + \frac{Q^{(5)}_p}{r_p} \, ,\nonumber\\
K^{p} &~=~ k^p_\infty + \frac{a_p \, \kappa_p}{q \, r_{p}} \, ,  \qquad  \qquad &&L_{p} &&~=~  \frac{Q^{(p)}_1}{r_1} +\frac{Q^{(p)}_5}{r_5} \, ,\nonumber
\end{alignat}
We have defined on purpose the ``effective" dipole charges $\kappa_I$ as a function of the charges in $K^I$: $\kappa_I = \frac{q \,k_I}{a_I}$. Those effective dipole charges have been argued to be the local magnetic charges obtained by integrating the magnetic dipole strength $da_I$ around the center \cite{Bena:2009en,Vasilakis:2011ki}.
Using the expression of the warp factors \eqref{eq2:Zgeneralform} and the general expressions for $\mu$ and $\omega$ in the \cite{Bena:2009en}, we obtain for our specific solutions
\begin{alignat}{2}
Z_1 &~=~ &&  \frac{Q^{(1)}_5}{r_5} + \frac{Q^{(1)}_p}{r_p} + \frac{\kappa_5 \kappa_p}{q} \frac{r}{r_5 r_p}\, ,\nonumber \\
Z_5 &~=~ && \frac{Q^{(5)}_1}{r_1} + \frac{Q^{(5)}_p}{r_p}  +\frac{\kappa_1 \kappa_p}{q} \frac{r}{r_1 r_p}\, , \nonumber\\
Z_p&~=~ &&  \frac{Q^{(p)}_1}{r_1} +\frac{Q^{(p)}_5}{r_5} + \frac{\kappa_1 \kappa_5}{q} \frac{r}{r_1 r_5}\, , \qquad\qquad \qquad \qquad \qquad\qquad \qquad \qquad\qquad \quad\label{eq3:3supZmuomega}\\
\mu &~=~ && \sum_I \sum_{J\neq I} \frac{Q_J^{(I)}\kappa_I}{2\, q}\, \frac{r^2 +a_I a_J - 2 a_I r \cos\theta}{(a_J-a_I) r_I r_J} + \frac{\kappa_1 \kappa_5 \kappa_p}{q^2} \, \frac{r^2 \cos \theta}{r_1 r_5 r_p} + \frac{r\,M}{q}\,, \nonumber
\end{alignat}
\begin{alignat}{2}
\omega &~=~ && \left[ \sum_I \sum_{J\neq I} \frac{Q_J^{(I)}\kappa_I}{2}\, \frac{r (a_J + a_I \cos2\theta) -(r^2 +a_I a_J)\cos \theta}{(a_J-a_I) r_I r_J} + \frac{\kappa_1 \kappa_5 \kappa_p}{q} \, \frac{r^2 \sin^2 \theta}{r_1 r_5 r_p}  \right. \nonumber \\
& && \left. \, +\, \omega_0 - \sum_I m_I \cos \theta_I - m_0 \cos \theta \right] \,d\phi \,,\nonumber
\end{alignat}
where we have defined the polar angles $\theta_I$ which correspond to the angle with the z-axis with respect to the center $I$
\begin{equation}
\cos\theta_I ~\equiv~ \frac{r \cos \theta -a_I}{r_I}.
\end{equation}
In order to analyze the spectrally-flowed solutions, we need to compute the electromagnetic gauge fields $A_I$  of the initial solutions
\begin{equation}
\begin{split}
w_I &~=~  \left( \kappa_I \,\frac{r-a_I \,\cos\theta}{r_I}- q\, k^I_\infty\, \cos\theta \right)d\phi\,, \\
A_I &~=~ -\frac{dt+\mu (d\psi+q\cos\theta \,d\phi) +\omega }{Z_I} + K_I (d\psi+q\cos\theta \,d\phi) +w_I\,.
\end{split}
\label{eq3:3supAw}
\end{equation}
For the same reason, the electromagnetic one-forms $v_0$ and $v_I$ involved in the spectral flow transformations of the gauge fields must be derived. This has not been done yet in the literature. We solve their equations \eqref{eq3:v0vIequations} in the context of our solutions in the appendix \ref{app:almostBPSgeneral}\footnote{This result can be easily generalized to generic multicenter almost-BPS solutions.}:
\begin{alignat}{2}
v_I &~=~ \frac{|\epsilon_{IJK}|}{2} &&\left[ -\,q\, k^J_\infty  k^K_\infty~ \cos \theta  \,+\, 2\, k^J_\infty \kappa^K ~ t_J^{(2)} \,-\, 2\,Q_J^{(I)} ~ t_J^{(1)} \,-\, \frac{\kappa_J \kappa_K}{q}~ t_{JK}^{(4)}\: \right]\, d\phi \,,\nonumber\\
v_0 &~=~ \frac{|\epsilon_{IJK}|}{6} &&\,\bigg[ q\,k^I_\infty  k^J_\infty  k^K_\infty~ \cos \theta  \,+\, 6\, k^I_\infty Q_J^{(I)}\, t_J^{(1)} \,-\, 3\, k^I_\infty k^J_\infty \kappa_K \, t_K^{(2)} \label{eq3:3supvv0}\\ 
& &&~\,+\, 6\,  \frac{\kappa_I Q_J^{(I)}}{q} \, t_{IJ}^{(3)} \,+\, 3 \, \frac{k^I_\infty \kappa_J \kappa_K}{q}\,t_{JK}^{(4)} \,+\, \frac{\kappa_I \kappa_J \kappa_K}{q^2} \, t_{IJK}^{(5)} \:\bigg] \, d\phi \,, \nonumber
\end{alignat}
with
\begin{alignat}{1}
t^{(1)}_I &~\equiv~ \cos \theta_I  \,, \qquad \qquad \quad  t^{(3)}_{IJ} ~\equiv~ \frac{a_I}{a_J-a_I} \, \frac{r^2 + a_I a_J -(a_I+a_J)r\cos \theta }{r_I r_J}\,, \nonumber\\
t^{(2)}_I &~\equiv~\frac{r-a_I \cos\theta}{r_I} \,, \qquad ~\, t ^{(4)}_{IJ} ~\equiv~  \frac{(r^2 + a_I a_J) \cos\theta -(a_I+a_J)r }{r_I r_J}\,,\\
t^{(5)}_{IJK} &~\equiv~ \frac{r^3 + r (a_I a_J+a_I a_K+a_J a_K)- \left(r^2(a_I + a_J+ a_K) +a_I a_J a_K \right) \cos \theta}{r_I r_J r_K}\,.\nonumber
\end{alignat}
At this point, we have the full description of the almost-BPS solutions we will use as input for our construction. In general, most of the solutions in this class are not regular. We investigate the regularity conditions in the next section.

\subsubsection{The regularity conditions and conditions on the asymptotics}
\label{sec:regularity3S}

\begin{itemize}
\item 16-supercharge regular two-charge supertube:
\end{itemize}
\noindent
A single two-charge supertube with dipole charge corresponding to, say, $K_p$ gives a regular six-dimensional metric if its angular-momentum charge is fixed to be $m_p = \frac{q \,Q_p^{(1)}Q_p^{(5)}}{2 a_p \kappa_p}$ \cite{Bena:2008dw,Bena:2009en}. Imposing such a condition to the three types of supertubes, we obtain the usual supertube regularity 
\begin{equation}
m_I ~=~ \frac{|\epsilon_{IJK}|}{2} \,\frac{q\, Q_I^{(J)}Q_I^{(K)}}{2 a_I \kappa_I} \, , \qquad I=1,5,p.
\label{eq3:superregcond}
\end{equation}

\begin{itemize}
\item Absence of Dirac-Misner strings at the centers:
\end{itemize}
\noindent
The absence of Dirac-Misner string singularities in $\omega$ requires $\omega|_{\theta=0,\pi}=0$. From \eqref{eq3:3supZmuomega}, this gives one condition on the constant term in $\omega$ and four bubble equations
\begin{equation}
\begin{split}
\omega_0 &~=~ 0\,\\
2 \,m_0 &~=~  \frac{\Gamma_{15}}{a_1-a_5} + \frac{\Gamma_{1p}}{a_1-a_p} + \frac{\Gamma_{5p}}{a_5-a_p}  \,, \\
\frac{q\, Q_1^{(5)} Q_1^{(p)}}{a_1 \kappa_1} &~=~ \frac{\Gamma_{15}}{|a_1-a_5|} + \frac{\Gamma_{1p}}{|a_1-a_p|} \,,\\
\frac{q\, Q_5^{(1)} Q_5^{(p)}}{a_5 \kappa_5} &~=~ \frac{\Gamma_{51}}{|a_1-a_5|} + \frac{\Gamma_{5p}}{|a_5-a_p|} \,,\\
\frac{q\, Q_p^{(1)} Q_p^{(5)}}{a_p \kappa_p} &~=~ \frac{\Gamma_{p1}}{|a_1-a_p|} + \frac{\Gamma_{p5}}{|a_5-a_p|} \,, 
\end{split}
\label{eq3:BE3supertubes}
\end{equation}
where $\Gamma_{IJ} \equiv \kappa_I Q_J^{(I)}- \kappa_J Q_I^{(J)}$.

\begin{itemize}
\item Absence of closed timelike curves:
\end{itemize}
The absence of closed timelike curves in the rest of the space requires the positivity of the quartic invariant $\mathcal{I}_4$ \eqref{eq2:I4}. This condition is in general very complicated to check directly since it is not an algebraic condition. However, it has been showed in \cite{Heidmann:2017cxt} that it can be systematically satisfied for our configurations. We just review briefly here the mechanism. A necessary condition for $\mathcal{I}_4 >0$ is to have 
\begin{equation}
Z_I V\,>\,0 \,, \quad I=1,5,p\,, \qquad \mu \underset{r \rightarrow\infty}{\rightarrow} 0.
\label{eq3:necCTCcondition}
\end{equation}
Imposing $ \mu\rightarrow0$ requires straightforwardly that $m_\infty ~=~0$. Furthermore, if we expand $Z_I V$ around the poles which is sufficient to prove \eqref{eq3:necCTCcondition} we get the following conditions:
\begin{equation}
q\frac{Q_J^{(I)}}{a_J} + \frac{\kappa_I \kappa_J}{|a_I -a_J|} > 0 \qquad \text{and} \qquad q\left(\frac{Q_J^{(I)}}{a_J } + \frac{Q_K^{(I)}}{a_K }\right) > 0\, , \qquad I \neq J \neq K .
\label{eq3:I4positivitySimpler}
\end{equation}
This is trivially solved by taking all the supertube charges and $q$ to be positive. However, if one sums the three last bubble equations \eqref{eq3:BE3supertubes}, at least one supertube dipole charge needs to be negative. Let us consider only one negative charge, say $\kappa_5$. The conditions \eqref{eq3:I4positivitySimpler} will just define a significantly large domain of possible values. 

A priori, the condition \eqref{eq3:necCTCcondition} only guarantees that $Z_1 Z_5 Z_pV > 0$ which does not necessarily mean $\mathcal{I}_4>0$. However, from the construction of $\mu$ \eqref{eq2:EOM}, this is practically always sufficient. Moreover, we want our initial solutions to satisfy $j_L= j_R = 0$ as explained in section \ref{sec:asymptotics}. So $\mu$ decays as $r^{-2}$ rather than $r^{-1}$ which gives a stronger evidence for this fact.

\begin{itemize}
\item Conditions on the asymptotics
\end{itemize}
Our solutions need to be asymptotic to the specific S$^1$ fibration over AdS$_2$ giving AdS$_3$ \eqref{eq2:targetAdSmetric}. For that purpose, the constant terms in the warp factors $Z_I$ and in the Gibbons-Hawking function $V$ have been set to zero at the beginning \eqref{eq3:3supZmuomega}. Furthermore, to obtain the specific S$^1$ fibration, the right and left angular momenta must be zero as detailed in section \ref{sec:asymptotics}. They can be derived from the asymptotic behaviour of $\mu$
\begin{equation}
\mu  ~\underset{r \rightarrow\infty}{\sim}~ \frac{j_R + j_L \cos\theta}{r}.
\end{equation}
The AdS$_2$ throat has an infinite length due to the vanishing constant terms. This means that $j_R=0$ is straightforwardly satisfied. We obtain $j_L$ from \eqref{eq3:3supZmuomega}:
\begin{equation}
j_L  ~=~  \frac{2}{q}\left(\frac{\kappa_1 \kappa_5 \kappa_p}{q} + \sum_{I\neq J\neq K} \frac{Q_I^{(J)} Q_I^{(K)}}{\kappa_I} + \frac{1}{2} \sum_{I\neq J} \kappa_I Q_J^{(I)}\right) = 0 .
\label{eq3:JR}
\end{equation}
Our initial almost-BPS solutions must satisfy this equation before applying the sequence of generalized spectral flows.

\bigskip
\bigskip

In this section, we have described in full detail the family of extremal non-supersymmetric three-supertube solutions with a flat $\mathbb{R}^4$ base space and with zero left and right angular momenta. We have shown a procedure to construct systematically bubbling solutions of this type. We expect from section \ref{sec:adstowads} that acting with three generalized spectral flows on those initial solutions will produce our expected smooth bubbling asymptotically WAdS$^3\times$SqS$^3$ or NHEK geometries. We will discuss this construction in the next section.

\section{Asymptotically WAdS$_3\times$SqS$^3$ bubbling geometries}
\label{sec:bubblingWAdS}

We start with the solutions constructed in the previous section. We will perform three generalized spectral flows parametrized by the constant shifts $\gamma_1$, $\gamma_5$ and $\gamma_p$. Even if the transition from an AdS$_3$ to a WAdS$_3$ with generalized spectral flows seems to be straightforward from the point of view of section \ref{sec:adstowads}, things get more complicated for a bubbling geometry and we will need to massage the initial solutions and the spectral flows to satisfy different regularity conditions in the UV and IR geometries:
\begin{itemize}
\item The spectrally flowed UV geometry differs from a WAdS$_3\times$SqS$^3$ geometry by the angle periodicities \eqref{eq1:NHEKperiodicities} even if we start with an initial solution which is asymptotic to the right S$^1$ fibration over AdS$_2$. In section \ref{sec:UVgeoWAdS}, we will deal with the spectral flow parameters and the parameters of the initial solution to get a UV geometry exactly identified as a WAdS$_3\times$SqS$^3$ geometry with the right angle periodicities \eqref{eq1:NHEKperiodicities}. 
\item The spectrally flowed IR geometry is a smooth bubbling geometry. However, the modification of the angle periodicities in the UV region changes drastically the periods around the centers. Conical singularities can occur at the centers where the S$^3$ shrinks. We will show in section \ref{sec:IRgeoWAdS} that one can still systematically build geometries where the UV angle periods do not yield to conical singularities. 
\end{itemize}

Several attempts on building bubbling geometries with a NHEK or WAdS$_3\times$SqS$^3$ region have been performed in the previous work \cite{Bena:2015pua,Bena:2012wc}. In \cite{Bena:2012wc}, only very specific WAdS$_3$ geometries with limited field contents have been built. Furthermore, in both papers, the NHEK regions were built in the deep IR and the issue of conical singularities which can occur at the centers was not tackled. Here we give all the details of the construction of the largest known family of smooth general solutions with a WAdS$_3\times$SqS$^3$ UV.

\subsection{The ultraviolet geometry}
\label{sec:UVgeoWAdS}
We start with a solution of the family of almost-BPS solutions (detailed in \ref{sec:3SinTNgeneralities}) with all the constraints and regularity conditions satisfied. Thus, the asymptotic behavior of the initial solution is
\begin{equation}
\begin{split}
Z_I ~\sim ~ \frac{q_I}{r} \,,\qquad  K^I ~\sim ~ k_\infty^I \,,\qquad  V ~\sim ~ \frac{q}{r} \, ,\qquad \mu = \omega = \mathcal{O}(r^{-2}),\qquad r \gg 1,
\end{split}
\label{eq4:initialasymptquant}
\end{equation}
By applying the spectral flow transformation rules \eqref{eq3:SFrules}, the solution after three generalized spectral flows has the following asymptotic expansion:
\begin{equation}
\begin{split}
&\widetilde{Z}_I ~\sim ~ \frac{\widetilde{q}_I}{r} \,,\qquad  \widetilde{V} ~\sim ~ \frac{\widetilde{q}}{r} \, ,\qquad \widetilde{\mu}  ~\sim ~ \frac{\widetilde{J}}{r}\, ,\qquad  \widetilde{W}_I ~\sim ~ \frac{\widetilde{\chi}_I}{r} \, ,\qquad  \widetilde{P}_I ~\sim ~ \widetilde{k}_\infty^I  \\
&\widetilde{A}~\sim ~ \left( \widetilde{A}^{(0)}_\infty + \widetilde{A}_\infty \cos \theta \right) d\phi \,,\qquad  \widetilde{w}_I~\sim ~ \left( \widetilde{w}^{(0)}_{I\infty} +\widetilde{w}_{I\infty}\cos \theta \right) d\phi \,,\qquad r\rightarrow \infty\,,
\end{split}
\label{eq4:afterSFasymptquant}
\end{equation}
where each tilded quantity in the right-hand side is a constant which can be derived from \eqref{eq3:SFrules} as a function of the asymptotic values of the initial solution  \eqref{eq4:initialasymptquant}. Since these functions are rather complicated and of minor interest, we did not write them down in their general forms. However, it is noteworthy that $\widetilde{q}$ is generically a square root of a polynomial function. In anticipation of the constraints demanded by the regularity around the centers, one needs to impose all the quantities to be at least rational. For that purpose, we fix the polynomial to be a perfect square. Two simple choices are: $\gamma_p = 0$ and $\gamma_p = -\frac{1}{k_\infty^p}$. We have analyzed both possibilities and it happens that the second one leads to simpler solutions. From now on, we suppose that $\gamma_p = -\frac{1}{k_\infty^p}$. We define the constants
\begin{equation}
t_\infty^I ~\equiv~ 1+k_\infty^I\gamma_I\,.
\end{equation}
Then, we have\footnote{We use $t_\infty^p = 1+k_\infty^p\gamma_p =0$.}
\begin{alignat}{3}
\widetilde{q} &~=~\left| \frac{q_1 \gamma_5 \, t_\infty^1  - q_5 \gamma_1 \, t_\infty^5}{k_\infty^p}\right|\,, \qquad && \widetilde{q}_I &&~=~ \frac{q \, q_I \, (t_\infty^I)^2 +\frac{|\epsilon_{IJK}|}{2} \gamma_I^2 q_J q_K}{\widetilde{q}} \,, \nonumber\\
\widetilde{J} &~=~ \frac{q_1 q_5}{k_\infty^p \,\widetilde{q}^2} \left(\gamma_1 \gamma_5 \, q_p + q\, t_\infty^1\, t_\infty^5 \right)\,,\qquad && \widetilde{\chi}_I && ~=~ |\epsilon_{IJK}|\: \frac{2\widetilde{q}\,\widetilde{q}_I}{\gamma_J \gamma_K t_\infty^I q_I - 2\,\gamma_I \gamma_J t_\infty^K q_K} \,, \nonumber\\
\widetilde{k}_\infty^I &~=~ \frac{q \, q_I \, k_\infty^I \, t_\infty^I +\frac{|\epsilon_{IJK}|}{2} \gamma_I q_J q_K}{q \, q_I \, (t_\infty^1)^2 +\frac{|\epsilon_{IJK}|}{2} \gamma_I^2 q_J q_K} + \frac{\widetilde{J}}{\widetilde{\chi}_I}\,, \qquad &&\widetilde{A}_\infty && ~=~ -\frac{q_1 \gamma_5 \, t_\infty^1  + q_5 \gamma_1 \, t_\infty^5}{k_\infty^p}, \label{eq4:afterSFcharges}\\
\widetilde{w}_{I\infty} &~=~ -\frac{|\epsilon_{IJK}|}{2} \left[\, q\, k_\infty^I t_\infty^J t_\infty^K + 2 \,q_J \gamma_K + \gamma_J \gamma_K  \right. &&\sum &&\left.   k_\infty^L q_L   \, \right].\nonumber
\end{alignat}
The expressions of $\widetilde{w}^{(0)}_{I\infty}$ and $\widetilde{A}^{(0)}_\infty$ remain complicated functions of the charges of the initial solution and the interested reader can easily compute  them from \eqref{eq3:3supAw} and \eqref{eq3:3supvv0}. One can check by curiosity that the asymptotic value of the quartic invariant $\mathcal{I}_{4\, \infty}$ \eqref{eq2:I4} is indeed preserved
\begin{equation}
\mathcal{I}_{4 \,\infty} ~=~ q \,q_1 q_5 q_p ~=~ \widetilde{q} \,\widetilde{q}_1 \widetilde{q}_5 \widetilde{q}_p -  \widetilde{q}^2 \widetilde{J}^2  ~=~ \widetilde{\mathcal{I}}_{4 \,\infty}.
\label{eq4:asymptoticI4}
\end{equation}

By inserting \eqref{eq4:afterSFasymptquant} in the spectrally flowed six-dimensional metric \eqref{eq3:JR}, the WAdS$_3\times$SqS$^3$ asymptotic expansion of the metric is explicit
\begin{equation}
\begin{split}
ds_{\infty}^2 ~=~ \frac{\kappa^2}{4} &\left[-r^2 d\tau^2 + \frac{dr^2}{r^2} \,+\, \gamma \,(dy_\infty +r\, d\tau)^2 + \gamma \, ( d\psi_\infty + \cos\theta  d\phi )^2   \right. \\
 &\left.   \:+\: 2\, \alpha \, (dy_\infty +r\, d\tau) ( d\psi_\infty + \cos\theta \, d\phi ) \,+\, d\theta^2 + \sin^2 \theta \,d\phi^2\right. \bigg] +\, \ldots\,,
\end{split}
\label{eq4:asymptmetric}
\end{equation}
where we have defined the six-dimensional coordinates at infinity $(\tau,r,\theta,\phi,\psi_\infty, y_\infty)$ using the initial coordinates $(t,r,\theta,\phi,\psi, y)$ as follows
\begin{equation}
\begin{split}
 y_\infty &~ \equiv ~\sqrt{\mathcal{I}_{4 \,\infty}} \,\widetilde{\chi}_p \,\frac{\widetilde{w}_{p\infty} \,\left(\psi \,+\,\widetilde{A}^{(0)}_{\infty}\, \phi \right) \:-\: \widetilde{A}_{\infty}\left(y \,+\,\widetilde{w}^{(0)}_{p\infty}\, \phi \right)}{\widetilde{A}_{\infty}\, \left(\mathcal{I}_{4 \,\infty} - \widetilde{q}^2  \widetilde{\chi}_p  \widetilde{k}_\infty^p \widetilde{J} \right)-  \widetilde{w}_{p\infty}  \: \widetilde{q}^2  \widetilde{\chi}_p  \widetilde{J} }\,, \qquad \tau ~ \equiv ~ \frac{t}{\sqrt{\mathcal{I}_{4 \,\infty}}}\,, \\
 \psi_\infty &~ \equiv ~ \frac{\left(\psi +\widetilde{A}^{(0)}_{\infty}\,\phi \right)\left(\mathcal{I}_{4 \,\infty} - \widetilde{q}^2  \widetilde{\chi}_p  \widetilde{k}_\infty^p \widetilde{J}\right) \:-\: \widetilde{q}^2  \widetilde{\chi}_p  \widetilde{J} \left(y +  \widetilde{w}^{(0)}_{p\infty} \,\phi\right)}{\widetilde{A}_{\infty} \left(\mathcal{I}_{4 \,\infty} - \widetilde{q}^2  \widetilde{\chi}_p  \widetilde{k}_\infty^p \widetilde{J} \right)-   \widetilde{w}_{p\infty}  \: \widetilde{q}^2  \widetilde{\chi}_p  \widetilde{J} }\,,
\label{eq4:WadSchangeofvar}
\end{split}
\end{equation}
and where the warp constant factors, $\gamma$ and $\alpha$, and the length, $\kappa$, are given by
\begin{equation}
\begin{split}
\gamma &~=~\left(\frac{\widetilde{q}_p}{\widetilde{w}_p}\right)^2 \,\left[\,1 \,+\,\frac{\widetilde{q} \,\widetilde{J}^2}{\widetilde{q}_1 \widetilde{q}_5 \widetilde{q}_p} \left( \left(\frac{\widetilde{w}_p}{\widetilde{q}_p}\right)^2-1\right)\right]\,,\\
\alpha &~=~ - \frac{\sqrt{ \mathcal{I}_{4 \,\infty}} }{\widetilde{q}^3 \,\widetilde{q}_1 \widetilde{q}_5 \widetilde{q}_p} \left({\widetilde{A}_{\infty}} \left(\widetilde{k}_\infty^p\, \widetilde{q}_p^2 + \widetilde{\chi}_p \widetilde{J} \right) \:+\: \widetilde{q}_p^2 \, \widetilde{w}_{p\infty} \right) \,,\\
\kappa^2 &~=~ 4 \,\widetilde{q} \sqrt{\widetilde{q}_1 \widetilde{q}_5}\,.
\end{split}
\label{eq4:alphgammawarpedfactor}
\end{equation}
The last condition to obtain WAdS$_3\times$SqS$^3$ in the asymptotic region is on the periods for $(y_\infty,\psi_\infty,\phi)$ 
\begin{equation}
(y_\infty,\psi_\infty,\phi) =
\left\{
\arraycolsep=1.6pt\def\arraystretch{1.6}
\begin{array}{rl}
(y_\infty,\psi_\infty,\phi) \,\,+\,\, &2\pi \,(T_y , -T_\psi  ,0)\\
(y_\infty,\psi_\infty,\phi) \,\,+\,\, & 2\pi \,(0, 2  ,0)\\
(y_\infty,\psi_\infty,\phi) \,\,+\,\, &2\pi \,(0 , 1  ,1)
\end{array}
\right. .
\label{eq4:WadSperiods}
\end{equation}
Such periodicities are complicated to obtain while keeping the usual periods for $(y,\psi,\phi)$
\begin{equation}
y =y +2\pi \,,\qquad \psi =\psi+4\pi \,,\qquad (\psi,\phi)=(\psi,\phi)+(2\pi,2\pi). 
\label{eq4:usualangleperiods}
\end{equation}
However, one can just reverse the perspective by imposing directly the periods \eqref{eq4:WadSperiods} for $(y_\infty,\psi_\infty,\phi)$ and express the corresponding periods of $(y,\psi,\phi)$ by inverting \eqref{eq4:WadSchangeofvar}. This has the advantage of adding no new complicated constraints on the parameters of the solution but the main drawback is that this drastic modification of periods of $(y,\psi,\phi)$ can induce conical singularities in the IR wherever the S$^3$ shrinks. 

\subsection{The infrared geometry}
\label{sec:IRgeoWAdS}

Generalized spectral flows preserve the bubbling feature of the initial solution: the number of centers and their positions on the $\mathbb{R}^3$ base space are straightforwardly preserved. They also preserve all the conditions for the absence of closed timelike curves as detailed in section \ref{sec:ABPSspectralflows}. Moreover, they transform a singular supertube center to a smooth Gibbons-Hawking center\footnote{A two-charge supertube is regular only when its magnetic dipole charge sources the P charge of the system. In our convention, this is a two-charge supertube of type p. Singular supertube centers are supertubes which source magnetically the D1 or the D5 charges \cite{Mateos:2001qs}.}. A series expansion of the spectrally flowed solution around the center $J$ (where $J=0,1,5,p$) gives
\begin{equation}
\begin{split}
&\widetilde{Z}_I ~\sim ~ \widetilde{z}_{IJ} \,,\qquad  \widetilde{V} ~\sim ~ \frac{\widetilde{q}_J}{r_J} \, ,\qquad \widetilde{\mu}  ~\sim ~ \widetilde{\mu}_J \,r_J\, ,\qquad  \widetilde{W}_I ~\sim ~ \widetilde{\chi}_{IJ} \, ,\qquad  \widetilde{P}_I ~\sim ~ \widetilde{k}_{IJ}  \\
&\widetilde{A}~\sim ~ \left( \widetilde{A}^{(0)}_J + \widetilde{A}_J \cos \theta_J \right) d\phi \,,\qquad \widetilde{w}_I~\sim ~ \left( \widetilde{w}^{(0)}_{IJ} +\widetilde{w}_{IJ}\cos \theta_J \right) d\phi \,,\qquad r_J \rightarrow 0\,,
\end{split}
\label{eq4:afterSFasymptquantcenter}
\end{equation}
where the tilded quantities in the right-hand sides are constant. It is not necessary for what will follow to write their complicated dependence on the parameters of the initial solution\footnote{For the interested reader, they can be easily derived with a calculation software using the transformation rules \eqref{eq3:SFrules} on the initial almost-BPS supertube solution given in \eqref{eq3:3supZmuomega}, \eqref{eq3:3supAw} and \eqref{eq3:3supvv0} and then taking the limit $r_J \rightarrow 0$.}. The three noteworthy points are
\begin{itemize}
\item[-] The ratio $\frac{\widetilde{A}_J}{\widetilde{q}_J}$ is equal to $1$. This is a key feature of an ambipolar Gibbons-Hawking metric. Indeed, in a generic Gibbons-Hawking metric \eqref{eq2:GHmetric}, the term proportional to $\cos\theta \, d\phi$ in $\frac{A}{V}$ must be exactly $\cos\theta\, d\phi$.
\item[-] The quantity $\widetilde{k}_{pJ}  \widetilde{A}_J + \widetilde{w}_{pJ}$ is equal to zero. Thus, the U(1) fiber defined by $\widetilde{A}_p + dy$ in \eqref{eq3:SFmetric} has no term proportional to $\cos\theta_J \,d\phi$ when we approach the center $J$. The local five-dimensional base space is then an exact direct product of a S$^1$ with a Gibbons-Hawking space.
\item[-] All the quantities in \eqref{eq4:afterSFasymptquantcenter} except $\widetilde{z}_{IJ}$ are rational functions of the initial parameters. This will be an important ingredient if we require the local geometry to be a discrete quotient of S$^1\times\mathbb{R}^4$.
\end{itemize}

\noindent
We can now use the expansions \eqref{eq4:afterSFasymptquantcenter} and the three remarks above to compute the limit of the spectrally flowed six-dimensional metric around the center $J$:
\begin{equation}
\begin{split}
ds_{J}^2 ~=~ \widetilde{q}_J \sqrt{\widetilde{z}_{1J}\widetilde{z}_{5J}}&\left[-d\tau_J^2 + \frac{dr_J^2}{r_J} \,+\, r_J \,\left(\left( d\psi_J +(1+\,\cos\theta_J)d\phi\right)^2 + d\theta_J^2 + \sin^2 \theta_J d\phi^2 \right) \right.\\
 &\left.  \:+\: \frac{\widetilde{z}_{pJ}}{\widetilde{q}_J\,\widetilde{z}_{1J}\widetilde{z}_{5J}} \,dy_J^2 \,\, \right],
\end{split}
\label{eq4:aroundcentermetric}
\end{equation}
where we have defined the six-dimensional Gibbons-Hawking coordinate system $(\tau_J,r_J,\theta_J, \psi_J , \phi ,y_J)$ as a function of the initial coordinates $(t,r,\theta,\phi,\psi, y)$:
\begin{equation}
\begin{split}
\tau_J &~\equiv~ \frac{t}{\sqrt{\widetilde{q}_J \, \widetilde{z}_{1J}\widetilde{z}_{5J}\widetilde{z}_{pJ}}} \, , \qquad r_J ~\equiv~ \sqrt{r^2 +a_J^2 - 2 a_J r \cos \theta}\, , \qquad \cos \theta_J ~\equiv~ \frac{r \cos \theta -a_J}{r_J}\, ,\\
\psi_J &~\equiv~ \frac{\psi + ( \widetilde{A}^{(0)}_J  -1)\,\phi}{\widetilde{q}_J}\,,\qquad y_J ~\equiv~ y \,-\, \frac{t}{\widetilde{\chi}_{pJ}} \,+\, \widetilde{k}_{pJ} \psi \,+\, \left(\widetilde{k}_{pJ} \,  \widetilde{A}^{(0)}_J + \widetilde{w}^{(0)}_{pJ}\right) \phi \,.
\end{split}
\label{eq4:changeofvarcenter}
\end{equation}
We recognize the metric of a  U(1) fiber over a Gibbons-Hawking space. Thus, the local geometry has no curvature singularity. However, a conical singularity can occur depending on the periodicities of $(y_J,\psi_J,\phi)$. If the periodicities were the usual Gibbons-Hawking periods \eqref{eq4:usualangleperiods}, the base space would be a discrete $\mathbb{Z}_{|\widetilde{q}_J|}$ quotient of S$^1\times\mathbb{R}^4$. The absence of conical singularity at $r_J=0$ would simply require that $\widetilde{q}_J$ is integer-valued and would impose some arithmetic constraints on the coefficients involved in \eqref{eq4:changeofvarcenter} \footnote{See \cite{Bena:2015drs} for examples of this kind.}. However, the modification of the periodicities at infinity have drastically modified the periods of $(y,\psi,\phi)$ and the smoothness analysis will require the full mathematical machinery which we briefly detail following \cite{Giusto:2012yz,Chakrabarty:2015foa,Bossard:2017vii}.

\bigskip
Let us first map the Gibbons-Hawking patch of angles $(\theta_J,\psi_J,\phi)$ to the S$^3$ patch $(\theta_J,\phi_{LJ},\phi_{RJ})$ by taking
\begin{equation}
\phi_{RJ} ~=~ \frac{\psi_J}{2}\,,\qquad \phi_{LJ} ~=~ \frac{\psi_J}{2} \,+\, \phi\,.
\label{eq4:changevartoR4}
\end{equation}
The spacelike components of the metric ($dt_J=0$) gives the spherically symmetric metric on S$^1\times\mathbb{R}^4$
\begin{equation}
\begin{split}
\frac{1}{ \widetilde{q}_J \sqrt{\widetilde{z}_{1J}\widetilde{z}_{5J}}}\,\,ds_{a}^2 ~=~ & \frac{dr_J^2}{r_J} \,+\, \frac{\widetilde{z}_{pJ}}{\widetilde{q}_J\,\widetilde{z}_{1J}\widetilde{z}_{5J}} \,dy_J^2 \, \\
 & \:+\: r_J\bigg[d\theta_J^2 +2 (1+ \cos \theta_J) \,d\phi_{LJ}^2+2 (1- \cos \theta_J) \,d\phi_{RJ}^2 \bigg] \,,
\end{split}
\label{eq4:aroundcentermetricR4S1}
\end{equation}
The periodicities of $(y_J,\phi_{LJ},\phi_{RJ})$ can be read off from the periodicities of $(y_{\infty},\psi_{\infty},\phi)$ \eqref{eq4:WadSperiods} with the sequence of three linear changes of coordinates $(y_{\infty},\psi_{\infty},\phi)\rightarrow (y,\psi,\phi)$ in \eqref{eq4:WadSchangeofvar}, $(y,\psi,\phi)\rightarrow (y_J,\psi_J,\phi)$ in \eqref{eq4:changeofvarcenter} and $(y_J,\psi_J,\phi)\rightarrow (y_J,\phi_{LJ},\phi_{RJ})$ in \eqref{eq4:changevartoR4}. After few lines of computation, the periodicities translate into the following identifications\footnote{For readability, we have dropped the index $J$ referring to the center but the coefficient $\alpha_i$ and $\beta_i$ are not identical for the four centers.}
\begin{equation}
(y_J,\phi_{LJ},\phi_{RJ}) =
\left\{
\arraycolsep=1.6pt\def\arraystretch{1.6}
\begin{array}{rl}
(y_J,\phi_{LJ},\phi_{RJ}) \,\,+\,\, &2\pi \,(\alpha_A , \beta_A  , \beta_A) \quad \qquad \,\,\,\, (A)\\
(y_J,\phi_{LJ},\phi_{RJ}) \,\,+\,\, & 2\pi \,(\alpha_B, \beta_B  , \beta_B)  \quad \qquad \,\,\,\,(B)\\
(y_J,\phi_{LJ},\phi_{RJ}) \,\,+\,\, &2\pi \,(\alpha_C , 1+ \beta_C  ,\beta_C)  \qquad (C)
\end{array}
\right. ,
\label{eq4:identificationsatcenter}
\end{equation}
where the coefficients $\alpha_i$ and $\beta_i$ are complicated but computable rational functions depending on:
\begin{itemize}[topsep=0pt]
 \setlength\itemsep{-0.4em}
\item[-] The parameters of the initial almost-BPS solution.
\item[-] The spectral flow parameters $\gamma_I$.
\item[-] The periods $T_y$ and $T_\psi$ of the angles of the UV WAdS$_3$ \eqref{eq1:NHEKperiodicities}.
\item[-] The square root of the asymptotic value of the quartic invariant $\sqrt{\mathcal{I}_{4 \,\infty}}$ \eqref{eq4:asymptoticI4}.
\end{itemize}
The local geometry is a discrete quotient of S$^1\times\mathbb{R}^4$ if $\alpha_J$ and $\beta_J$ are rational numbers. Thus, all the initial parameters and  $\sqrt{\mathcal{I}_{4 \,\infty}}$ must be rational\footnote{This means that the entropy of the corresponding black hole given by $S=2\pi\sqrt{\mathcal{I}_{4 \,\infty}}$ belongs to $2\pi \mathbb{Q}$.}. Choosing the other parameters to be rational is easy. However, imposing $\sqrt{\mathcal{I}_{4 \,\infty}}$ to be rational requires a little bit of arithmetic.

Conical singularities only occur at points that are invariant under the operation
\begin{equation}
A^{n_A} B^{n_B} C^{n_C}\,,\qquad (n_A,n_b,n_c) \in \mathbb{Z}\,.
\label{eq4:conicalsingoperation}
\end{equation}
Furthermore, they all arise at $r_J=0$ where $\phi_{LJ}$ and $\phi_{RJ}$ are both degenerate, at $\theta_J=0$ where $\phi_{RJ}$ is degenerate and at $\theta_J=\pi$ where $\phi_{LJ}$ is degenerate. The periods of $\phi_{RJ}$ and $\phi_{LJ}$ are almost identical with a difference of $2\pi$ for the periodicity $C$, so if the identifications \eqref{eq4:identificationsatcenter} at $r_J=0$ do not destroy smoothness, they will also ensure the absence of singularities at $\theta_J=0$ or $\pi$ .

At $r_J=0$, in order for the shifts $\phi_{LJ} \rightarrow \phi_{LJ} +2\pi$ and $\phi_{RJ} \rightarrow \phi_{RJ} +2\pi$ at fixed $y_J$ to be a closed orbit, any triplet of integers $(n_A,n_B,n_C)$ where $y_J \rightarrow y_J$ under \eqref{eq4:conicalsingoperation} must satisfy $n_A \beta_A +n_B \beta_B +n_C \beta_C \in \mathbb{Z}$. In more concrete terms, any operation \eqref{eq4:conicalsingoperation} which leaves $y_J$ invariant, that is to say where  $n_A \alpha_A +n_B \alpha_B +n_C \alpha_C =0$, must transform $\phi_{LJ} \rightarrow \phi_{LJ} +2\pi \, N$ and $\phi_{LJ} \rightarrow \phi_{LJ} +2\pi \, N'$ where $N$ and $N'$ are both integers. Using simple arithmetic arguments one can show that this is equivalent to prove the condition for the three sets of integers $(0,n_B,n_C)$, $(n_A,0,n_C)$ and $(n_A,n_B,0)$. If the conditions are satisfied for each set, the action of the quotient is free and the local geometry around the center $J$ is then a smooth discrete quotient of S$^1\times\mathbb{R}^4$.

This analysis applies at every center. The total number of smoothness conditions is then 12 (3$\times$4 centers). The number of parameters is still greater than the number of conditions which gives good hope to draw a systematic construction procedure.

\subsection{The construction procedure}
\label{sec:sysconsWAdS}

We sketch briefly a technical summary of what we have done until now to build asymptotically WAdS$_3\times$SqS$^3$ geometries:
\begin{itemize}
\item We start with the family of almost-BPS four-center solutions of three supertubes in $\mathbb{R}^4$. Initially, it is a family of 15 rational parameters : $q$, $Q_I^{(J)}$, $\kappa^I$, $a_I$ and $k^I_\infty$. The regularity of the solution imposes three bubble equations \eqref{eq3:BE3supertubes}, the condition on the asymptotics requires $j_R=0$ \eqref{eq3:JR} and the positivity of the quartic invariant $\mathcal{I}_4$ is satisfied by imposing all the initial charges and dipole charges to be positive except one. Furthermore, $\sqrt{\mathcal{I}_{4\infty}}$ needs to be a rational number which is not an equation but one can consider that this fixes a parameter. We have consequently a 10-parameter family of initial almost-BPS solutions. 
\item After three generalized spectral flows, we have three new parameters $\gamma_I$ whose one is fixed to have rational spectrally flowed charges. Moreover, the periods $T_y$ and $T_\psi$ of the angles of the WAdS$_3\times$SqS$^3$ region can also be considered as free parameters. 
\item We have in total a 14-parameter family of bubbling asymptotically WAdS$_3\times$SqS$^3$ geometries. The smoothness of the geometry in the IR requires 12 arithmetic conditions as discussed in the previous section. These conditions do not exactly fix parameters so the parameter space of the resulting family of smooth solutions is complicated to define. However, many solutions can be easily generated by generating parameters and by checking for each set of parameters if the 12 arithmetic conditions can be satisfied. We give an example of such a solution in the next section. 
\end{itemize}

\subsection{An explicit example}
\label{sec:exampleWAdS}

We construct an explicit example of the procedure discussed above. We picked an almost-BPS three-supertube solution in $\mathbb{R}^4$ giving the solution which we use in the first step of the procedure:
\begin{alignat}{8}
q&~=~ \Lambda\,,\qquad  &&\kappa_1 &&~=~ \frac{\Lambda}{2}\,,\qquad  &&\kappa_5 &&~=~ -\frac{2 \Lambda}{3}\,,\qquad  && \kappa_p &&~=~ \frac{\Lambda}{2}\,, \qquad  Q^{(1)}_5 &&~=~ \Lambda \,, \nonumber \\
Q^{(1)}_p &~=~ \frac{2 \Lambda}{3} \,,\qquad && Q^{(5)}_1 &&~=~ \Lambda\,,\qquad &&  Q^{(5)}_p &&~=~ \Lambda \,,\qquad  && Q^{(p)}_1 &&~=~ \frac{\Lambda}{3} \,, \qquad  Q^{(p)}_5 &&~=~ \frac{4 \Lambda}{3}\,, \nonumber \\
\end{alignat}
where $\Lambda \in \mathbb{Q}^+$ is a degree of freedom of the charges which does not compromise the regularity of the solution and the condition on the asymptotics. We can consider $\Lambda$ as a free parameter all along the construction. The coordinates of the three supertube centers on the z-axis are
\begin{equation}
a_1 ~=~ 1 \,, \qquad a_5 ~=~ \frac{36}{13} \,, \qquad  a_p ~=~ 24\,.
\label{eq4:centerpositionsex}
\end{equation}
The solution is asymptotically AdS$_2\times$S$^1\times$S$^3$ which implies that the center positions are scaling invariant $a_I \rightarrow \lambda a_I$ \cite{Bena:2018bbd}. Consequently, one can freely rescale \eqref{eq4:centerpositionsex} to make the inter-center distances as small as we want. 

\noindent We did not fix yet the constant terms $k_\infty^I$ since they are not involved in the regularity of the solution. They are actually irrelevant from the point of view of the initial almost-BPS solution since they affect only the asymptotic values of the gauge, $A_I$, of the solution which can be gauge-fixed to zero. However, they affect greatly the solutions one obtains after spectral flows.

From \eqref{eq3:3supZmuomega}, one can derive the asymptotic values of the D1, D5, P charges of the initial solution, the left and right angular momenta and the entropy of the corresponding three-charge black hole
\begin{equation}
\begin{split}
q_1 &~=~ \frac{4 \Lambda}{3} \,, \qquad q_5 ~=~ \frac{9 \Lambda}{4} \,, \qquad q_p ~=~ \frac{4 \Lambda}{3} \,,\\
j_L &~=~ j_R ~=~ 0 \,, \\
S &~=~ 2 \pi \sqrt{\mathcal{I}_{4\infty}} ~=~ 4\pi \Lambda^2.
\end{split}
\end{equation}

One can now play with the spectral flow parameters $\gamma_I$ and the constants $k_\infty^I$ to generate an extremal non-supersymmetric smooth asymptotically WAdS$_3\times$SqS$^3$ bubbling geometry. We found an infinite number of such solutions. To give an example, we pick one of these solutions:
\begin{equation}
\gamma_1 ~=~\frac{1}{2}\, ,\quad \gamma_5 ~=~-1\, ,\quad \gamma_p ~=~-1\, ,\quad k_\infty^1 ~=~-\frac{3}{2}\, ,\quad k_\infty^5 ~=~3\, ,\quad k_\infty^p ~=~1\,.
\end{equation}
We can derive the full geometry by computing the metric and the gauge fields \eqref{eq3:SFmetric} and \eqref{eq3:SFrules}.
We will just focus on the WAdS$_3\times$SqS$^3$ asymptotic region which  is given by
\begin{equation}
\begin{split}
ds_{\infty}^2 ~=~ \frac{1}{3}\sqrt{\frac{485}{6}} &\left[-r^2 d\tau^2 + \frac{dr^2}{r^2} \,+\, \frac{701}{485} \,(dy_\infty +r\, d\tau)^2 + \frac{701}{485}  \, ( d\psi_\infty + \cos\theta  d\phi )^2   \right. \\
 &\left.   \:+\: \frac{1302}{485} \, (dy_\infty +r\, d\tau) ( d\psi_\infty + \cos\theta \, d\phi ) \,+\, d\theta^2 + \sin^2 \theta \,d\phi^2\right. \bigg].
\end{split}
\label{eq4:asymptmetricex}
\end{equation}
 We choose the periodicities of the angles to be\footnote{Many other possibilities were available.}
\begin{equation}
(y_\infty,\psi_\infty,\phi) =
\left\{
\arraycolsep=1.6pt\def\arraystretch{1.6}
\begin{array}{rl}
(y_\infty,\psi_\infty,\phi) \,\,+\,\, &2\pi \,\left(\dfrac{24}{43} \,T , -\dfrac{8}{11}\, T  ,0\right)\\
(y_\infty,\psi_\infty,\phi) \,\,+\,\, & 2\pi \,(0, 2  ,0)\\
(y_\infty,\psi_\infty,\phi) \,\,+\,\, &2\pi \,(0 , 1  ,1)
\end{array}
\right. ,
\label{eq4:WadSperiodsex}
\end{equation}

where $T$ is a free parameter. For the reader interested in the smoothness of the bubbling geometry in the IR, the metric and the periodicities of the angles around the centers are given in the appendix \ref{app:solWadS}. We found that the IR bubbling geometry is smooth if and only if  $T=\frac{a}{b}\in\mathbb{Q}$ and $b$ is not divisible by 2 or  13.


\section{Asymptotically NHEK bubbling geometries }
\label{sec:bubblingNHEK}

In the previous section, we have constructed in detail a large family of extremal non-supersymmetric bubbling solutions which cap off smoothly in the IR and which are asymptotically WAdS$_3\times$SqS$^3$. In the present section, we push a bit further the construction to  asymptotically NHEK bubbling solutions. The path from WAdS$_3\times$SqS$^3$ to NHEK requires to relate the WAdS$_3$ region of our solutions to the near-horizon region of the over-rotating 5d D1-D5-P black hole  detailed in section \ref{sec:5DD1D5PBH}. This essentially means that we have to express the charges, angular momentum and mass of the D1-D5-P black hole in terms of the parameters of our solutions. Once this is done, we have to impose the NHEK periodicities of the angles at infinity \eqref{eq1:NHEKperiods} and check the smoothness of the IR geometry as it has been done for asymptotically WAdS$_3$ geometries in section \ref{sec:IRgeoWAdS}. At first sight, this might seem to be a mere formality. However, the fact that the periods of $y_\infty$ and $\psi_\infty$ were free parameters for asymptotically WAdS$_3$ solutions was practical to satisfy the twelve conditions of smoothness at the centers. Now that the periods are connected to the parameters defining the bubbling geometry, this requires more work.

In this section, we use all the results obtained in the previous section. We have started with the family of almost-BPS three-supertube solutions in $\mathbb{R}^4$ and performed the sequence of generalized spectral flows detailed in section \ref{sec:UVgeoWAdS} to obtain an asymptotically WAdS$_3\times$SqS$^3$ bubbling geometry. In section \ref{sec:UVgeoNHEK} we will match this asymptotic region to a near-horizon region of an extremal non-supersymmetric D1-D5-P black hole. We will identify the corresponding periodicities and see how such solutions can be systematically generated in section \ref{sec:sysconsNHEK}. At the end of the section, we will give an explicit example of a solution.

\subsection{Matching the WAdS$_3$ UV geometry to NHEK}
\label{sec:UVgeoNHEK}

After applying the sequence of generalized spectral flows to our family of almost-BPS solutions, the asymptotic metric is given by \eqref{eq4:asymptmetric} where the constant warp factors $\gamma$ and $\alpha$ and the length $\kappa$ are defined in \eqref{eq4:alphgammawarpedfactor}. We want to relate this geometry to the near-horizon geometry of an extremal non-supersymmetric D1-D5-P black hole determined by four parameters $a$, $\delta_1$, $\delta_5$ and $\delta_p$ and given by the metric \eqref{eq1:NHEKWAdSmetric} where $\gamma$, $\alpha$ and $\kappa$ are defined by \eqref{eq1:WAdSfactors}. We use the three identities between $\gamma$, $\alpha$ and $\kappa$ to relate $a$, $\delta_1$ and $\delta_5$ to the parameters of our solutions and we use the matching of the entropy to find $\delta_p$. After few lines of computation, we obtain
\begin{equation}
\begin{split}
a &~=~ 2 \left(\mathcal{I}_{4\infty} \,\gamma_1\, \gamma_5\, t_\infty^1\, t_\infty^5  \right)^{1/4}\,, \\
s_1 &~=~ \frac{1}{2}\left[ \frac{q\, q_1 \,{t_\infty^1}^2 + \gamma_1 \left(\gamma_1 \,q_5 \,q_p - 2 \sqrt{\mathcal{I}_{4\infty}}\, t_\infty^1 \right)}{ \gamma_1 \,t_\infty^1 \sqrt{\mathcal{I}_{4\infty}}}\right]^{1/2}\,,\\
s_5 &~=~ \frac{1}{2}\left[ \frac{q\, q_5 \,{t_\infty^5}^2 + \gamma_5 \left(\gamma_5 \,q_1 \,q_p - 2 \sqrt{\mathcal{I}_{4\infty}}\, t_\infty^5 \right)}{ \gamma_5 \,t_\infty^5 \sqrt{\mathcal{I}_{4\infty}}}\right]^{1/2}\,,\\
s_p &~=~ \frac{\mathcal{I}_{4\infty}}{4 \,a^3}\,\frac{s_1 \,s_5 + c_1 \,c_5 \,\sqrt{\mathcal{H}}}{c_1^2\,c_5^2 -s_1^2\,s_5^2} \,,
\end{split}
\label{eq5:pathtoNHEK}
\end{equation}
where $s_I = \sinh \delta_I$, $c_I = \cosh \delta_I$ and $\mathcal{H}$ is defined as
\begin{equation}
\begin{split}
\mathcal{H} &~\equiv~ 1 - 16\, \frac{a^6 \left(c_1^2\,c_5^2 -s_1^2\,s_5^2 \right)}{\mathcal{I}_{4\infty}}\\
& ~=~ 1 - 256\, \sqrt{\gamma_1 \,\gamma_5 \,t_\infty^1 \,t_\infty^5 }\,\left(\gamma_1 \,q_5\, t_\infty^5 + \gamma_5 \,q_1\, t_\infty^1\right)\,\left(q\, t_\infty^1 \,t_\infty^5 + q_p\,\gamma_1 \, \gamma_5 \right).
\label{eq5:H}
\end{split}
\end{equation}
The mass, the D1, D5 and P charges and the left angular momentum of the corresponding Cvetic-Youm black hole can be derived using \eqref{eq1:D1D5PchargesmassAM}. However, the most interesting quantities are the NHEK periods $T_y$ and $T_\psi$ \eqref{eq1:NHEKperiods}. Using \eqref{eq5:pathtoNHEK}, we can show that they are rational numbers if:
\begin{itemize}
\item[-] $\sqrt{\mathcal{I}_{4\infty}}$ is rational. This is the same condition as the one imposed for asymptotically WAdS$_3$ bubbling geometries.
\item[-] $\sqrt{\mathcal{H}}$ is rational. This is a more complicated condition to satisfy than the previous one. Tricky arithmetic is required. 
\end{itemize}

Once the matching to NHEK is performed, one can look at the IR bubbling region of our solutions. The periodicities of the angles around each center depends on $T_y$ and $T_\psi$. Thus the local geometries are quotients of $\mathbb{R}^4\times$S$^1$ only if they are rational. Moreover, conical singularities might still occur as for asymptotically WAdS$_3$ bubbling geometries. We use the same smoothness analysis as in section \ref{sec:UVgeoWAdS} to derive 12 conditions to have smooth discrete quotients on $\mathbb{R}^4\times$S$^1$ around the centers.

\subsection{The construction procedure}
\label{sec:sysconsNHEK}
The procedure to build smooth asymptotically NHEK bubbling geometries is similar to the one depicted in section \ref{sec:sysconsWAdS}:
\begin{itemize}
\item We start with the family of almost-BPS four-center solutions of three supertubes in $\mathbb{R}^4$. Initially, it is a family parametrized by 15 rational parameters : $q$, $Q_I^{(J)}$, $\kappa^I$, $a_I$ and $k^I_\infty$. We solve the three bubble equations \eqref{eq3:BE3supertubes}, $j_R=0$ \eqref{eq3:JR} and we require the positivity of the quartic invariant $\mathcal{I}_4$ by imposing all the initial charges to be positive except one. Furthermore, $\sqrt{\mathcal{I}_{4\infty}}$ needs to be a rational number which fixes a parameter. We have consequently a 10-parameter family of initial almost-BPS solutions. 
\item After three generalized spectral flows, we have two more parameters $\gamma_I$($\gamma_p$ is fixed to have rational spectrally flowed charges). The condition on the periods $T_y$ and $T_\psi$ to be rational requires some arithmetic machinery which fixes 3 parameters (the two remaining spectral flow parameters and $k_\infty^5$).
\item We have in total an 8-parameter family of bubbling asymptotically NHEK$_3$ geometries. The smoothness of the geometry in the IR requires 12 arithmetic conditions. Even if the parameter space is not easy to determine, we can perform a loop generating technique to build a large number of such solutions. We give an example of such a solution in the next section. 
\end{itemize}

\subsection{An explicit example}
\label{sec:exampleNHEK}

We construct an explicit example of the procedure discussed above. We choose a smooth almost-BPS three-supertube solution in $\mathbb{R}^4$ satisfying the first point of the procedure:
\begin{alignat}{8}
q&~=~1\,,\qquad  &&\kappa_1 &&~=~\Lambda \,,\qquad  &&\kappa_5 &&~=~ -\frac{\Lambda}{2}\,,\qquad  && \kappa_p &&~=~ \frac{\Lambda}{3}\,, \qquad  Q^{(1)}_5 &&~=~\Lambda^2 \,, \nonumber \\
Q^{(1)}_p &~=~ \frac{\Lambda^2}{2} \,,\qquad && Q^{(5)}_1 &&~=~ \frac{2 \Lambda^2}{3}\,,\qquad &&  Q^{(5)}_p &&~=~ \frac{\Lambda^2}{3} \,,\qquad  && Q^{(p)}_1 &&~=~ \frac{\Lambda^2}{2} \,, \qquad  Q^{(p)}_5 &&~=~ \Lambda^2\,, \nonumber \\
\end{alignat}
where $\Lambda \in \mathbb{Q}^+$ corresponds to the charge-scaling free parameter. We choose a slightly different charge-scaling $\Lambda$ than in section \ref{sec:exampleWAdS}. They are actually equivalent. The present choice is just more adapted to the matching with NHEK. The coordinates of the three supertube centers on the z-axis are
\begin{equation}
a_1 ~=~ 1 \,, \qquad a_5 ~=~\frac{3}{14}\left(17 + \sqrt{65}\right)\,, \qquad  a_p ~=~ \frac{3}{4}\left(9 + \sqrt{65}\right)\,.
\label{eq5:centerpositionsex}
\end{equation}
Once again, the AdS$_2\times$S$_1\times$S$^3$ asymptotics of the solution allows us to rescale $a_I \rightarrow \lambda a_I$ as small as we want. The irrationality of the inter-center distances does not impact the smoothness of the solution around the centers.

From \eqref{eq3:3supZmuomega}, one can derive the asymptotic values of the initial D1, D5, P charges of the solution, the left and right angular momenta and the entropy of the initial system
\begin{equation}
\begin{split}
q_1 &~=~ \frac{4 \Lambda^2}{3} \,, \qquad q_5 ~=~ \frac{4 \Lambda^2}{3} \,, \qquad q_p ~=~ \Lambda^2 \,,\\
j_L &~=~ j_R ~=~ 0 \,, \\
S &~=~ 2 \pi \sqrt{\mathcal{I}_{4\infty}} ~=~ \frac{8\pi}{3} \Lambda^3.
\end{split}
\end{equation}

One can now play with the spectral flow parameters $\gamma_I$ and the constants $k_\infty^I$ to generate an extremal non-supersymmetric smooth asymptotically NHEK$_3\times$SqS$^3$ bubbling geometry. This requires $\sqrt{\mathcal{H}}$ in \eqref{eq5:H} to be rational. After a rather technical arithmetic computation we found several values for $\gamma_1$, $\gamma_5$ and $k_\infty^5$ which lead to rational NHEK periods $T_y$ and $T_\psi$ without inducing any conical singularities at the centers:
\begin{alignat}{5}
\gamma_1 &~=~\frac{1}{3\,\Lambda-2}\, ,\qquad &&\gamma_5 &&~=~\frac{9\,\Lambda -6}{40 \,\Lambda^2}\, ,\qquad &&\gamma_p &&~=~-1\, ,\nonumber \\
k_\infty^1 &~=~2\, ,\qquad &&k_\infty^5 &&~=~-\frac{\Lambda}{3}\,\frac{2+37\, \Lambda}{3 \,\Lambda -2}\, ,\qquad &&k_\infty^p &&~=~1\,.
\end{alignat}
We can derive the full geometry by computing the metric and the gauge fields \eqref{eq3:SFmetric} and \eqref{eq3:SFrules}.
We will just focus on the NHEK asymptotic region which is given by
\begin{equation}
\begin{split}
ds_{\infty}^2 ~=~ \frac{\Lambda^2}{3} &\left[-r^2 d\tau^2 + \frac{dr^2}{r^2} \,+\, \frac{34}{25} \,(dy_\infty +r\, d\tau)^2 + \frac{34}{25} \, ( d\psi_\infty + \cos\theta  d\phi )^2   \right. \\
 &\left.   \:+\: \frac{12}{5} \, (dy_\infty +r\, d\tau) ( d\psi_\infty + \cos\theta \, d\phi ) \,+\, d\theta^2 + \sin^2 \theta \,d\phi^2\right. \bigg].
\end{split}
\label{eq5:asymptmetric}
\end{equation}
\noindent The NHEK angle periodicities are
\begin{equation}
(y_\infty,\psi_\infty,\phi) =
\left\{
\arraycolsep=1.6pt\def\arraystretch{1.7}
\begin{array}{rl}
(y_\infty,\psi_\infty,\phi) \,\,+\,\, &2\pi \,\left(\dfrac{75}{32 \,\Lambda} , -\dfrac{93}{32\,\Lambda}  ,0\right)\\
(y_\infty,\psi_\infty,\phi) \,\,+\,\, & 2\pi \,(0, 2  ,0)\\
(y_\infty,\psi_\infty,\phi) \,\,+\,\, &2\pi \,(0 , 1  ,1)
\end{array}
\right. .
\label{eq5:NHEKperiodsex}
\end{equation}
The asymptotic NHEK region corresponds to the near-horizon region of an extremal Kerr black hole given by the following mass, angular momenta and charges:
\begin{equation}
\begin{split}
M ~&=~ \frac{287\, \Lambda^2}{75}, \\
J_R ~&=~ 0\,,\qquad J_L ~=~\frac{124 \Lambda^3}{75}, \\
Q_1~&=~ \frac{16 \Lambda^2}{15} \,,\qquad Q_5~=~ \frac{16 \Lambda^2}{15}\,,\qquad Q_p~=~ \frac{21 \Lambda^2}{25}  
\end{split}
\end{equation}

For the reader interested in the feature of the bubbling geometry in the IR, we gave the local metrics and the periodicities of the angles around the centers in the appendix \ref{app:solNHEK}. 

For any rational values of $\Lambda$, we found a smooth non-supersymmetric extremal geometry which is bubbling in the IR and NHEK in the UV. 

One can also be interested in computing the RR three-form flux $F^{(3)}$ of the final solution. One can perform a similar computation of the spectrally flowed RR three-form flux $F^{(3)}$ as in \cite{Bena:2015pua}.


\section{Conclusions}
\label{sec:conclusion}

In this paper, we have constructed a family of smooth bubbling solutions in six dimensions which are asymptotic to either generic WAdS$_3\times$SqS$^3$ or NHEK. We gave explicit examples of the construction which can be used for different purposes:
\begin{itemize}
\item One can investigate their CFT dual states. They can give some hints on the nature of the CFT$_2$ dual to WAdS$_3$ or the CFT$_2$ dual to NHEK.
\item Nearly extreme black hole have been seen in the sky \cite{McClintock:2006xd}. From an astrophysical point of view, one can compute the Kerr multipole moments of our solutions to see if there exist deviations from the Kerr-Newman black hole solution. This could give interesting observable quantities in order to detect some imprints of the microstate structure of black holes in the gravitational wave emission after a  collision of two black holes.
\end{itemize}

Furthermore, one can extend the construction to have microstates of the whole Kerr-Newman black hole solution in five dimensions and not only its near-horizon geometry. In the context of multicenter solutions, we have to ``bring back the ones" in the harmonic functions which will make the solutions to be asymptotically flat. This is the subject of future work.


\section*{Acknowledgments}

I am grateful to Iosif Bena, Guillaume Bossard and Monica Guica for their useful advice. The work of PH was supported by a CDSN from ENS Lyon and by the ANR grant Black-dS-String ANR-16-
CE31-0004-01. 

\appendix

\section{Derivation of $v_I$ and $v_0$}
\label{app:almostBPSgeneral}

In this section, we derive the electromagnetic one-forms $v_0$ and $v_I$. They are involved in the generalized spectral flow transformations of an almost-BPS multicenter solutions  \eqref{eq3:SFrules}. This is why they have to be derived in order to match the asymptotics of the spectrally flowed solutions to WAdS$_3$ or NHEK and to regularize them. They satisfy the equations \eqref{eq3:v0vIequations} where $Z_I$, $K_I$ and $V$ are given \eqref{eq3:initialsupsol} and \eqref{eq3:3supZmuomega}. We decompose the equations as follows 
\begin{alignat}{2}
\star_3 d v_I &~=~ \frac{|\epsilon_{IJK}|}{2} &&\bigg[ -\,q\, k^J_\infty  k^K_\infty~ \star_3 d T^{(0)}  \,+\, 2\, k^J_\infty \kappa^K ~ \star_3 d  T_J^{(2)} \,-\, 2\,Q_J^{(I)} ~ \star_3 d  T_J^{(1)}\nonumber \\
& && ~\,-\, \frac{\kappa_J \kappa_K}{q}~ \star_3 d T_{JK}^{(4)}\: \bigg] \,,\\
\star_3 d v_0 &~=~ \frac{|\epsilon_{IJK}|}{6} &&\,\bigg[ q\,k^I_\infty  k^J_\infty  k^K_\infty~ \star_3 d T^{(0)}  \,+\, 6\, k^I_\infty Q_J^{(I)}\, \star_3 d T_J^{(1)} \,-\, 3\, k^I_\infty k^J_\infty \kappa_K \, \star_3 d  T_K^{(2)} \nonumber\\ 
& &&~\,+\, 6\,  \frac{\kappa_I Q_J^{(I)}}{q} \, \star_3 d T_{IJ}^{(3)} \,+\, 3 \, \frac{k^I_\infty \kappa_J \kappa_K}{q}\,\star_3 d T_{JK}^{(4)} \,+\, \frac{\kappa_I \kappa_J \kappa_K}{q^2} \, \star_3 d T_{IJK}^{(5)} \:\bigg]\,, \nonumber
\end{alignat}
where $T^{(0)}$, $T_I^{(2)}$, $T_{IJ}^{(3)}$, $T_{IJ}^{(4)}$ and $T_{IJK}^{(5)} $ satisfy
\begin{alignat}{2}
\star_3 d T^{(0)}  &~=~&& d\left(\frac{1}{r}\right)\,,\qquad \star_3 d T_I^{(1)}~=~ d\left(\frac{1}{r_I}\right) \,,\qquad \star_3 d T_I^{(2)}~=~ \frac{a_I}{r} d\left(\frac{1}{r_I}\right) - \frac{a_I}{r_I} d\left(\frac{1}{r}\right) \,,\nonumber \\
\star_3 d T_{IJ}^{(3)} &~=~&&  \frac{a_I}{r_I} d\left(\frac{1}{r_J}\right) - \frac{a_I}{r_J} d\left(\frac{1}{r_I}\right)\,,\qquad \star_3 d T_{IJ}^{(4)} ~=~  \left(1-\frac{a_I a_J}{r^2} \right) d\left(\frac{r}{r_I r_J} \right)\,,\\
\star_3 d T_{IJK}^{(5)} &~=~&& \left(\frac{1}{a_I a_J} +\frac{1}{r^2}- \frac{1}{a_I a_K}- \frac{1}{a_Ja_K}\ \right) \,\frac{r}{r_I r_J} d\left(\frac{1}{r_K}\right) \,\nonumber\\
& && \,~ + \left(\frac{1}{a_I a_K} +\frac{1}{r^2}- \frac{1}{a_I a_J}- \frac{1}{a_J a_K}\ \right) \,\frac{r}{r_I r_K} d\left(\frac{1}{r_J}\right) \,\nonumber\\
& && \,~ + \left(\frac{1}{a_J a_K} +\frac{1}{r^2}- \frac{1}{a_I a_J}- \frac{1}{a_I a_K}\ \right) \,\frac{r}{r_J r_K} d\left(\frac{1}{r_I}\right) \,\nonumber\\
& && \,~ + \left(-\frac{1}{r^2} +\frac{1}{a_I a_J} + \frac{1}{a_I a_K}+ \frac{1}{a_J a_K}\ \right) \,\frac{r^2}{r_I r_K r_K} d\left(\frac{1}{r}\right) \,.\nonumber\\
\end{alignat}
We find
\begin{alignat}{1}
T^{(0)} &~\equiv~ \cos \theta \, d\phi  \,, \qquad  T^{(1)}_I ~\equiv~ \cos \theta_I \, d\phi  \,, \qquad T^{(2)}_I ~\equiv~\frac{r-a_I \cos\theta}{r_I} \,d\phi\,, \nonumber\\
T^{(3)}_{IJ} &~\equiv~ \frac{a_I}{a_J-a_I} \, \frac{r^2 + a_I a_J -(a_I+a_J)r\cos \theta }{r_I r_J}\,d\phi\,, \\
 T^{(4)}_{IJ} &~\equiv~  \frac{(r^2 + a_I a_J) \cos\theta -(a_I+a_J)r }{r_I r_J}\,d\phi\,,\nonumber\\
T^{(5)}_{IJK} &~\equiv~ \frac{r^3 + r (a_I a_J+a_I a_K+a_J a_K)- \left(r^2(a_I + a_J+ a_K) +a_I a_J a_K \right) \cos \theta}{r_I r_J r_K}\,d\phi\,.\nonumber
\end{alignat}
Thus, $v_0$ and $v_I$ are given by
\begin{alignat}{2}
 v_I &~=~ \frac{|\epsilon_{IJK}|}{2} &&\bigg[ -\,q\, k^J_\infty  k^K_\infty~ T^{(0)}  \,+\, 2\, k^J_\infty \kappa^K ~ T_J^{(1)} \,-\, 2\,Q_J^{(I)} ~ T_J^{(2)}\nonumber \\
& && ~\,-\, \frac{\kappa_J \kappa_K}{q}~ T_{JK}^{(3)}\: \bigg] \,,\\
v_0 &~=~ \frac{|\epsilon_{IJK}|}{6} &&\,\bigg[ q\,k^I_\infty  k^J_\infty  k^K_\infty~ T^{(0)}  \,+\, 6\, k^I_\infty Q_J^{(I)}\, T_J^{(1)} \,-\, 3\, k^I_\infty k^J_\infty \kappa_K \,  T_K^{(2)} \nonumber\\ 
& &&~\,+\, 6\,  \frac{\kappa_I Q_J^{(I)}}{q} \, T_{IJ}^{(3)} \,+\, 3 \, \frac{k^I_\infty \kappa_J \kappa_K}{q}\, T_{JK}^{(4)} \,+\, \frac{\kappa_I \kappa_J \kappa_K}{q^2} \,  T_{IJK}^{(5)} \:\bigg]\,. \nonumber
\end{alignat}

\section{The explicit asymptotically WAdS$_3\times$SqS$^3$ bubbling solution}
\label{app:solWadS}

In this section, we focus on the IR geometry of the solution constructed in section \ref{sec:exampleWAdS}. We give the local S$^1\times\mathbb{R}^{4}$ metric \eqref{eq4:aroundcentermetricR4S1} at each of the four centers and the periodicities of the angles.

\begin{itemize}
\item At the origin of the space, $r\sim 0$:
\end{itemize}
The local metric is
\begin{equation}
\begin{split}
ds_{0}^2 ~=~ \frac{5 \,\Lambda^2}{18}\sqrt{\frac{7}{3}}& \left[ \frac{dr^2}{r} \,+\, \frac{5632}{1575}\,dy_0^2 \, \right.\\
 & \left. \:+\: r\left(d\theta^2 +2 (1+ \cos \theta) \,d\phi_{L0}^2+2 (1- \cos \theta) \,d\phi_{R0}^2 \right)\right] \,,
\end{split}
\end{equation}
where $y_0$, $\phi_{L0}$ and $\phi_{R0}$ are related to the angles at infinity $y_\infty$, $\psi_{\infty}$ and $\phi$ by
\begin{equation}
\begin{split}
y_0 &~=~ y_\infty +\frac{9}{4}\left(\psi_\infty - \phi \right)\,,\\
\phi_{L0} &~=~ \frac{1}{8} \left(21 \, y_\infty + 31 ( \psi_\infty - \phi) \right) + \phi\,,\\
\phi_{R0} &~=~ \frac{1}{8} \left(21 \, y_\infty + 31 ( \psi_\infty - \phi) \right) \,.
\end{split}
\end{equation}
We can read the periodicities from \eqref{eq4:WadSperiodsex}
\begin{equation}
(y_0,\phi_{L0},\phi_{R0}) =
\left\{
\arraycolsep=1.6pt\def\arraystretch{2}
\begin{array}{rl}
(y_0,\phi_{L0},\phi_{R0}) \,\,-\,\, & \dfrac{20\,\pi\,T}{473}\,\left(51, 64 ,64\right)\\
(y_0,\phi_{L0},\phi_{R0}) \,\,+\,\, & \dfrac{\pi}{2} \,\left(18, 31 , 31\right)\\
(y_0,\phi_{L0},\phi_{R0}) \,\,+\,\, &2\pi \,\left(0 , 1 ,0\right)
\end{array}
\right. .
\end{equation}
Using the procedure detailed in section \ref{sec:IRgeoWAdS}, this corresponds to a smooth discrete quotient of S$^1\times\mathbb{R}^{4}$ if $b$ is not divisible by 2 where $b$ is the denominator of the irreducible fraction $T=\frac{a}{b}$.

\begin{itemize}
\item At the second center, $r_1 \sim 0$:
\end{itemize}
The local metric is
\begin{equation}
\begin{split}
ds_{1}^2 ~=~ \frac{ \Lambda^2}{23}\sqrt{\frac{155}{2}}& \left[ \frac{dr_1^2}{r_1} \,+\, \frac{15548}{837}\,dy_1^2 \, \right.\\
 & \left. \:+\: r_1\left(d\theta_1^2 +2 (1+ \cos \theta_1) \,d\phi_{L1}^2+2 (1- \cos \theta_1) \,d\phi_{R1}^2 \right)\right] \,,
\end{split}
\end{equation}
where $y_1$, $\phi_{L1}$ and $\phi_{R1}$ are related to the angles at infinity $y_\infty$, $\psi_{\infty}$ and $\phi$ by
\begin{equation}
\begin{split}
y_1 &~=~ y_\infty + \psi_\infty - \frac{5}{13} \phi \,,\\
\phi_{L1} &~=~ \frac{1}{6} \left(21 \, y_\infty + 31 \,\psi_\infty - 23\,\phi \right) + \phi\,,\\
\phi_{R1} &~=~\frac{1}{6} \left(21 \, y_\infty + 31 \,\psi_\infty - 23\,\phi \right)  \,.
\end{split}
\end{equation}
We can read the periodicities from \eqref{eq4:WadSperiodsex}
\begin{equation}
(y_1,\phi_{L1},\phi_{R1}) =
\left\{
\arraycolsep=1.6pt\def\arraystretch{2}
\begin{array}{rl}
(y_1,\phi_{L1},\phi_{R1}) \,\,-\,\, & \dfrac{160\,\pi\,T}{1419}\,\left(3, 32 ,32\right)\\
(y_1,\phi_{L1},\phi_{R1}) \,\,+\,\, & \dfrac{2\pi}{3} \,\left(6, 31 , 31\right)\\
(y_1,\phi_{L1},\phi_{R1})\,\,+\,\, &\dfrac{2 \pi}{39} \,\left(24 , 91 ,52\right)
\end{array}
\right. .
\end{equation}
This corresponds to a smooth discrete quotient of S$^1\times\mathbb{R}^{4}$ if $b$ is not divisible by 13 where $b$ is the denominator of the irreducible fraction $T=\frac{a}{b}$.

\begin{itemize}
\item At the third center, $r_5 \sim 0$:
\end{itemize}
The local metric is
\begin{equation}
\begin{split}
ds_{5}^2 ~=~ \frac{65\, \Lambda^2}{207}\sqrt{\frac{11}{6}}& \left[ \frac{dr_5^2}{r_5} \,+\, \frac{599081}{128700}\,dy_5^2 \, \right.\\
 & \left. \:+\: r_5\left(d\theta_5^2 +2 (1+ \cos \theta_5) \,d\phi_{L5}^2+2 (1- \cos \theta_5) \,d\phi_{R5}^2 \right)\right] \,,
\end{split}
\end{equation}
where $y_5$, $\phi_{L5}$ and $\phi_{R5}$ are related to the angles at infinity $y_\infty$, $\psi_{\infty}$ and $\phi$ by
\begin{equation}
\begin{split}
y_5 &~=~ y_\infty - \frac{3}{61}\left(3\,\psi_\infty - 13\, \phi \right)\,,\\
\phi_{L5} &~=~ \frac{1}{32} \left(21 \, y_\infty + 31 \,\psi_\infty - 17\,\phi \right) + \phi\,,\\
\phi_{R5} &~=~\frac{1}{32} \left(21 \, y_\infty + 31 \,\psi_\infty - 17\,\phi \right)  \,.
\end{split}
\end{equation}
Then, we can read the periodicities from \eqref{eq4:WadSperiodsex}
\begin{equation}
(y_5,\phi_{L5},\phi_{R5}) =
\left\{
\arraycolsep=1.6pt\def\arraystretch{2}
\begin{array}{rl}
(y_5,\phi_{L5},\phi_{R5})   \,\,+\,\, & \dfrac{320\,\pi\,T}{28853}\,\left(120, -61 ,-61\right)\\
(y_5,\phi_{L5},\phi_{R5})  \,\,+\,\, & \dfrac{\pi}{488} \,\left(-288, 1891 ,1891\right)\\
(y_5,\phi_{L5},\phi_{R5})  \,\,+\,\, &\dfrac{\pi}{488} \,\left(480 , 1403 ,427\right)
\end{array}
\right. .
\end{equation}
This corresponds to a smooth discrete quotient of S$^1\times\mathbb{R}^{4}$ if $b$ is not divisible by 16 where $b$ is the denominator of the irreducible fraction $T=\frac{a}{b}$.

\begin{itemize}
\item At the fourth center, $r_p \sim 0$:
\end{itemize}
The local metric is
\begin{equation}
\begin{split}
ds_{p}^2 ~=~ \frac{ \Lambda^2}{207}\sqrt{\frac{145}{3}}& \left[ \frac{dr_p^2}{r_p} \,+\, \frac{66309}{232}\,dy_p^2 \, \right.\\
 & \left. \:+\: r_p\left(d\theta_p^2 +2 (1+ \cos \theta_p) \,d\phi_{Lp}^2+2 (1- \cos \theta_p) \,d\phi_{Rp}^2 \right)\right] \,,
\end{split}
\end{equation}
where $y_p$, $\phi_{Lp}$ and $\phi_{Rp}$ are related to the angles at infinity $y_\infty$, $\psi_{\infty}$ and $\phi$ by
\begin{equation}
\begin{split}
y_p &~=~ y_\infty - \frac{21}{31}\left(\psi_\infty + \phi \right)\,,\\
\phi_{Lp} &~=~ \frac{1}{16} \left(21 \, y_\infty + 31 \,\psi_\infty + 15\,\phi \right) + \phi\,,\\
\phi_{Rp} &~=~\frac{1}{16} \left(21 \, y_\infty + 31 \,\psi_\infty +15 \,\phi \right)  \,.
\end{split}
\end{equation}
Then, we can read the periodicities from \eqref{eq4:WadSperiodsex}
\begin{equation}
(y_p,\phi_{Lp},\phi_{Rp}) =
\left\{
\arraycolsep=1.6pt\def\arraystretch{2}
\begin{array}{rl}
(y_p,\phi_{Lp},\phi_{Rp})    \,\,+\,\, & \dfrac{640\,\pi\,T}{14663}\,\left(3, -31 ,-31\right)\\
(y_p,\phi_{Lp},\phi_{Rp})   \,\,+\,\, & \dfrac{\pi}{124} \,\left(336, 961, 961\right)\\
(y_p,\phi_{Lp},\phi_{Rp})   \,\,+\,\, &\dfrac{\pi}{124} \,\left(336, 961, 713\right)
\end{array}
\right. .
\end{equation}
This corresponds to a smooth discrete quotient of S$^1\times\mathbb{R}^{4}$ if $b$ is not divisible by 8 where $b$ is the denominator of the irreducible fraction $T=\frac{a}{b}$.

Consequently, the IR bubbling geometry is smooth if and only if  $T=\frac{a}{b}\in\mathbb{Q}$ and $b$ is not divisible by 2 or  13.

\section{The explicit asymptotically NHEK bubbling solution}
\label{app:solNHEK}

In this section, we focus on the IR geometry of the solution constructed in \ref{sec:exampleWAdS}. We give the local S$^1\times\mathbb{R}^{4}$ metrics \eqref{eq4:aroundcentermetricR4S1} around each of the four centers and the periodicities of the angles.

\begin{itemize}
\item At the origin of the space, $r\sim 0$:
\end{itemize}
The local metric is
\begin{equation}
\begin{split}
ds_{0}^2 ~=~ \frac{\Lambda^2}{240}\sqrt{\frac{675-67\sqrt{65}}{6}}& \left[ \frac{dr^2}{r} \,+\, \frac{49 \,(2915 + 259 \sqrt{65})}{10240}\,dy_0^2 \, \right.\\
 & \left. \:+\: r\left(d\theta^2 +2 (1+ \cos \theta) \,d\phi_{L0}^2+2 (1- \cos \theta) \,d\phi_{R0}^2 \right)\right] \,,
\end{split}
\end{equation}
where $y_0$, $\phi_{L0}$ and $\phi_{R0}$ are related to the angles at infinity $y_\infty$, $\psi_{\infty}$ and $\phi$ by
\begin{equation}
\begin{split}
y_0 &~=~  y_\infty +\frac{17}{7}\left(\psi_\infty - \phi \right) \,,\\
\phi_{L0} &~=~ -6 \, y_\infty + 10\, ( \phi - \psi_\infty )  + \phi\,,\\
\phi_{R0} &~=~-6 \, y_\infty + 10 \,( \phi - \psi_\infty ) \,.
\end{split}
\end{equation}
The periodicities derived from \eqref{eq5:NHEKperiodsex} are
\begin{equation}
(y_0,\phi_{L0},\phi_{R0}) =
\left\{
\arraycolsep=1.6pt\def\arraystretch{2}
\begin{array}{rl}
(y_0,\phi_{L0},\phi_{R0}) \,\,+\,\, & 2\pi\,\left(-\dfrac{33}{7\,\Lambda}, \dfrac{15}{\Lambda} ,\dfrac{15}{\Lambda} ,\right)\\
(y_0,\phi_{L0},\phi_{R0}) \,\,+\,\, & 2\pi \,\left(\dfrac{34}{7}, -20 ,-20\right)\\
(y_0,\phi_{L0},\phi_{R0}) \,\,+\,\, &2\pi \,\left(0 , 1 ,0\right)
\end{array}
\right. .
\end{equation}
Using the procedure in section \ref{sec:IRgeoWAdS}, this is a smooth discrete quotient of S$^1\times\mathbb{R}^{4}$ for any rational $\Lambda$.

\begin{itemize}
\item At the second center, $r_1 \sim 0$:
\end{itemize}
The local metric is
\begin{equation}
\begin{split}
ds_{1}^2 ~=~ \frac{\Lambda^2}{420}\sqrt{\frac{-635+123\sqrt{65}}{2}} & \left[ \frac{dr_1^2}{r_1} \,+\,\frac{289 \,(2575+ 303 \sqrt{65})}{47360}\,dy_1^2 \, \right.\\
 & \left. \:+\: r_1\left(d\theta_1^2 +2 (1+ \cos \theta_1) \,d\phi_{L1}^2+2 (1- \cos \theta_1) \,d\phi_{R1}^2 \right)\right] \,,
\end{split}
\end{equation}
where $y_1$, $\phi_{L1}$ and $\phi_{R1}$ are related to the angles at infinity $y_\infty$, $\psi_{\infty}$ and $\phi$ by
\begin{equation}
\begin{split}
y_1 &~=~ y_\infty + \frac{1}{17}\,( 23\, \psi_\infty - 27\, \phi )\,,\\
\phi_{L1} &~=~ -6 \, y_\infty - 10 \,\psi_\infty + 12 \,\phi\,,\\
\phi_{R1} &~=~ -6 \, y_\infty - 10 \,\psi_\infty + 11 \,\phi  \,.
\end{split}
\end{equation}
The periodicities are
\begin{equation}
(y_1,\phi_{L1},\phi_{R1}) =
\left\{
\arraycolsep=1.6pt\def\arraystretch{2}
\begin{array}{rl}
(y_1,\phi_{L1},\phi_{R1}) \,\,-\,\, & 2\pi\,\left(-\dfrac{27}{17\,\Lambda}, \dfrac{15}{\Lambda} ,\dfrac{15}{\Lambda} ,\right)\\
(y_1,\phi_{L1},\phi_{R1}) \,\,+\,\, &2\pi \,\left(\dfrac{46}{17}, -20 ,-20\right)\\
(y_1,\phi_{L1},\phi_{R1})\,\,+\,\, &2\pi \,\left(-\dfrac{4}{17}, 2 ,1\right)
\end{array}
\right. .
\end{equation}
This is a smooth discrete quotient of S$^1\times\mathbb{R}^{4}$.

\begin{itemize}
\item At the third center, $r_5 \sim 0$:
\end{itemize}
The local metric is
\begin{equation}
\begin{split}
ds_{5}^2 ~=~\frac{\Lambda^2}{560}\sqrt{\frac{9635-323\sqrt{65}}{6}}  & \left[ \frac{dr_5^2}{r_5} \,+\,\frac{529 \,(1215+ 223 \sqrt{65})}{564480} \,dy_5^2 \, \right.\\
 & \left. \:+\: r_5\left(d\theta_5^2 +2 (1+ \cos \theta_5) \,d\phi_{L5}^2+2 (1- \cos \theta_5) \,d\phi_{R5}^2 \right)\right] \,,
\end{split}
\end{equation}
where $y_5$, $\phi_{L5}$ and $\phi_{R5}$ are related to the angles at infinity $y_\infty$, $\psi_{\infty}$ and $\phi$ by
\begin{equation}
\begin{split}
y_5 &~=~ y_\infty + \frac{1}{23}\left(\psi_\infty + 3\, \phi \right)\,,\\
\phi_{L5} &~=~ - \frac{1}{21} \left(6 \, y_\infty + 10 \,\psi_\infty + 9 \,\phi \right) + \phi\,,\\
\phi_{R5} &~=~ - \frac{1}{21} \left(6 \, y_\infty + 10 \,\psi_\infty + 9 \,\phi \right)  \,,
\end{split}
\end{equation}
with the following periodicities
\begin{equation}
(y_5,\phi_{L5},\phi_{R5}) =
\left\{
\arraycolsep=1.6pt\def\arraystretch{2}
\begin{array}{rl}
(y_5,\phi_{L5},\phi_{R5})   \,\,+\,\, & 2\,\pi \,\left(\dfrac{51}{23 \,\Lambda}, \dfrac{5}{7 \,\Lambda} ,\dfrac{5}{7 \,\Lambda}\right)\\
(y_5,\phi_{L5},\phi_{R5})  \,\,+\,\, &2\,\pi\,\left(\dfrac{2}{23}, -\dfrac{20}{21} , -\dfrac{20}{21}\right)\\
(y_5,\phi_{L5},\phi_{R5})  \,\,+\,\, &2\,\pi\,\left(\dfrac{4}{23}, \dfrac{2}{21} , -\dfrac{19}{21}\right)
\end{array}
\right. .
\end{equation}
This corresponds to a smooth discrete quotient of S$^1\times\mathbb{R}^{4}$.

\begin{itemize}
\item At the fourth center, $r_p \sim 0$:
\end{itemize}
The local metric is
\begin{equation}
\begin{split}
ds_{p}^2 ~=~ \frac{ 4\,\Lambda^2}{9}\frac{1}{43675+5435\sqrt{65}}& \left[ \frac{dr_p^2}{r_p} \,+\, \frac{45 \,(6145+ 737 \sqrt{65})}{8} \,dy_p^2 \, \right.\\
 & \left. \:+\: r_p\left(d\theta_p^2 +2 (1+ \cos \theta_p) \,d\phi_{Lp}^2+2 (1- \cos \theta_p) \,d\phi_{Rp}^2 \right)\right] \,,
\end{split}
\end{equation}
where $y_p$, $\phi_{Lp}$ and $\phi_{Rp}$ are related to the angles at infinity $y_\infty$, $\psi_{\infty}$ and $\phi$ by
\begin{equation}
\begin{split}
y_p &~=~ y_\infty + \frac{3}{5}\left(\psi_\infty + \phi \right)\,,\\
\phi_{Lp} &~=~ -6 \, y_\infty - 10 \,\psi_\infty - 9\,\phi\,,\\
\phi_{Rp} &~=~ -6 \, y_\infty - 10 \,\psi_\infty - 10\,\phi \,.
\end{split}
\end{equation}
The periodicities are
\begin{equation}
(y_p,\phi_{Lp},\phi_{Rp}) =
\left\{
\arraycolsep=1.6pt\def\arraystretch{2}
\begin{array}{rl}
(y_p,\phi_{Lp},\phi_{Rp})  \,\,-\,\, & 2\pi\,\left(\dfrac{3}{5\,\Lambda}, \dfrac{15}{\Lambda} ,\dfrac{15}{\Lambda} ,\right)\\
(y_p,\phi_{Lp},\phi_{Rp})  \,\,+\,\, &2\pi \,\left(\dfrac{6}{5}, -20 ,-20\right)\\
(y_p,\phi_{Lp},\phi_{Rp}) \,\,+\,\, &2\pi \,\left(\dfrac{6}{5}, -19 ,-20\right)
\end{array}
\right. .
\end{equation}
This corresponds to a smooth discrete quotient of S$^1\times\mathbb{R}^{4}$.

Consequently, the IR bubbling geometry is smooth for any rational values of $\Lambda$.

\newpage


\bibliography{references}
\bibliographystyle{utphysmodb}

\end{document}